\documentclass[12pt,draftclsnofoot,peerreview,onecolumn,notitlepage]{IEEEtran}
\IEEEoverridecommandlockouts

\usepackage{cite}
\usepackage{ieeefig}
\usepackage{amsfonts}
\usepackage{epsfig}
\usepackage{graphicx}
\usepackage[caption=false,font=footnotesize]{subfig}
\usepackage{stfloats}
\usepackage{amsmath}
\usepackage{color}

\newtheorem{theorem}{Theorem}
\newtheorem{lemma}{Lemma}

\newtheorem{remark}{Remark}

\newtheorem{definition}{Definition}
\newcommand{\define}    {\stackrel{\triangle}{=}}  

\begin{document}
%
\title{\LARGE{The Degrees of Freedom Regions of MIMO Broadcast, Interference, %
                  and Cognitive Radio Channels with No CSIT} }

\author{Chinmay S.~Vaze %
        and~Mahesh K.~Varanasi
\thanks{This work was supported in part by NSF Grant 0728955. The authors
are with the Department of Electrical, Computer, and Energy
Engineering, University of Colorado, Boulder, CO 80309-0425 USA
(e-mail: {Chinmay.Vaze, varanasi}@colorado.edu). The material in
this paper was presented in part at the IEEE Intl.
Symp. of Inform. Th., Austin, TX, Jun. 2010.} }

%



\maketitle

\begin{abstract}
The degrees of freedom (DoF) regions are characterized for the
multiple-input multiple-output (MIMO) broadcast channel (BC),
interference channels (IC) (including X and multi-hop interference
channels) and the cognitive radio channel (CRC),
when there is perfect and no channel state information at the
receivers and the transmitter(s) (CSIR and CSIT), respectively. For
the K-user MIMO BC, the exact characterization of the DoF region is
obtained, which shows that a simple time-division-based transmission
scheme is DoF-region optimal. Using the techniques developed for
the MIMO BC, the corresponding problems
for the two-user MIMO IC and the two-user MIMO CRC with an arbitrary
number of antennas at each of the four terminals are addressed.
For both of these channels, inner and outer bounds to the DoF
region are obtained and are seen to coincide for a vast
majority of the relative numbers of antennas at the four
terminals, thereby characterizing DoF regions for all but a few cases.
Finally, the DoF regions of the $K$-user MIMO IC, the CRC, and X networks are
derived for certain classes of these networks, including the one
where all transmitters have an equal number of antennas and
so do all receivers. The results of this paper are derived for
distributions of fading channel matrices and additive noises
that are more general than those considered in other simultaneous
related works. The DoF regions with and without CSIT are compared and conditions
on the relative numbers of antennas at the terminals under which a lack of CSIT does, or
does not, result in the loss of DoF are identified, thereby providing, on the one hand, simple and robust communication
schemes that don't require CSIT but have the same DoF performance as their previously found CSIT counterparts,
and on the other hand, identifying situations where CSI feedback to transmitters would provide gains that are significant
enough that even the DoF performance could be improved.
\end{abstract}




\section{Introduction}
\IEEEPARstart{M}ultiple-input multiple-output (MIMO) systems
are of great interest because they can provide a significantly
higher capacity as compared to their single-input single-output
(SISO) counterparts by exploiting the spatial dimension. One way of
measuring this benefit at high signal-to-noise ratio (SNR)
is via the spatial multiplexing gain or the
degrees of freedom (DoF), which is defined as the limit of the ratio
of the capacity to the logarithm of the SNR. For
example, the point-to-point MIMO channel with $M$ transmit and
$N$ receive antennas has $\min(M,N)$ DoF whereas its SISO
counterpart has only $1$ DoF \cite{Telatar}. Interestingly,
$\min(M,N)$ DoF are achievable over the MIMO channel even if there is
perfect channel state information (CSI) just at the receiver
(CSIR). In other words, the presence or absence of CSI at the
transmitter (CSIT) does not affect the DoF of the MIMO channel.
However, this may not necessarily be the case with multi-user
networks. Consider, for instance, two of the simplest multi-user
MIMO channels, namely, the multi-access channel and the broadcast
channel (BC). While the DoF region of the former is again not affected by
partial (or lack of) CSIT \cite{Telatar}, imperfect CSIT can severely impact the
DoF region of the BC \cite{Shamai-W-S, Caire, Jafar-Goldsmith,
Lapidoth, Chiachi}. With this motivation, we aim to comprehensively study the effect
of lack of CSIT on the DoF regions of several wireless MIMO networks including
the $K$-user BC, the 2-user interference \cite{Carleial, Jafar-Maralle, Chiachi-Jafar} and cognitive
radio channels (CRC) \cite{Chiachi-Jafar, Devroye, WeiWu, Sridharan} with an arbitrary number of
antennas at each of the four nodes, as well as certain classes of $K$-user interference, X and cognitive networks
(cf. \cite{Cadambe,JaferShamai, Khandani_Xchannel_constant,Jafar_Cadambe_wireless_X_network}) and multi-hop interference networks (cf. \cite{Cao_Chen_two_hop_IC_2009,Jafar_2hop_IC_DoF}).

The loss of DoF of a wireless channel due to no CSIT was
demonstrated for the first time in \cite{Caire} in the context of
the Gaussian MISO BC. Subsequently, for the BC with $M$ transmit
antennas, $K$ single-antenna users, and isotropic fading, it was
proved that the maximum sum-DoF achievable without CSIT is $1$,
which is significantly less than $\min(M,K)$ sum-DoF that are
attainable with perfect CSIT \cite{Jafar-Goldsmith}. In
\cite{Lapidoth}, the authors studied the real-valued Gaussian BC
with $2$ transmit antennas, $2$ single-antenna users, and any
arbitrary type of partial CSIT. They upper-bounded the achievable
sum-DoF by $\frac{2}{3}$ whereas $1$ DoF is achievable with CSIT.
This result, in spite of being available only in a special case with
its tightness unknown, is strong because it tells us that no matter
how good the quality of partial CSIT is, as long as it is not
perfect, the partial-CSIT sum-DoF can be significantly less than the
perfect-CSIT sum-DoF. Lastly, in \cite{Chiachi, Chiachi2}, the
authors studied the DoF of the two-user Gaussian MIMO BC under independent and
identically distributed (i.i.d.) Rayleigh fading and proved that the
no-CSIT DoF region can be exhausted by a simple time-division-based
scheme that transmits to only one user at a time.

In this paper, we derive the DoF region of the $K$-user MIMO BC (see
Theorems \ref{thm: dof region BC}--\ref{thm:D2}) under certain
general assumptions on fading distributions (that include the i.i.d.
Rayleigh fading model). It is proved that the DoF region of the
MIMO BC can be achieved by a simple strategy of time sharing. To
establish this result, the maximum weighted sum DoF 
achievable by time-division is shown to also be an outer bound to
the DoF region. Toward this end, the capacity region of the BC is
first outer-bounded by assuming that each receiver has genie-aided
knowledge of some of the messages that are not intended for it.
Under this assumption, the rate achievable for a given user is
upper-bounded, through Fano's inequality \cite{CT}, by the mutual
information (MI) between the signal received by that user and its
intended message, conditioned on the unintended messages its
receiver is assumed to know. These bounds thus imply that the above
weighted sum of the DoF is upper-bounded by the corresponding
weighted sum of the multiplexing gains of the MI terms. This latter
upper bound on the weighted sum is then shown to coincide with that
achievable with time-division by making an appropriate choice for
the genie-aided side-information. Moreover, the same result is shown
to be applicable for a wide class of distributions of channel
matrices and additive noises, including i.i.d.
Rayleigh fading, Rician fading, correlated Rayleigh
fading, non-Gaussian additive noise, and isotropic fading models as
well as the case where the channel matrices are correlated across
time. This makes our result more general than the previous results
of \cite{Jafar-Goldsmith, Chiachi, Chiachi2} and the no-CSIT special
case of \cite{Lapidoth} (see Remark \ref{rem: comparison BC with
other results}).

Next, we address the problem of characterizing the DoF region of the
no-CSIT MIMO IC. The two-user IC is first studied and an inner-bound
to its DoF region (see Theorem \ref{thm: inner bound IC}),
based on the basic techniques of time-sharing and receive
zero-forcing, is obtained. Clearly, this inner-bound can not involve the CSI-dependent
scheme of transmit zero-forcing beam-forming which is necessary
for DoF-region-optimality in the two-user MIMO IC with CSIT
\cite{Jafar-Maralle,Chiachi-Jafar}. This implies
that any signal stream that is intended for one receiver would cause
interference at the other. The receivers, being equipped with
perfect channel knowledge, can zero-force the interference to
recover the useful signal. Therefore, the inner-bound with no-CSIT
is in general smaller (but not always strictly) than the perfect-CSIT DoF region. Next, we
obtain an outer-bound to the DoF region (see Theorem \ref{thm: outer
bound IC}). To this end, the bounding technique developed while
solving the corresponding problem for the BC is used. The derived
inner and outer bounds are seen to coincide for a vast majority of
the values of the number of antennas at the four terminals. In particular,
for the MIMO IC in which the transmitters have
$M_1$ and $M_2$ antennas and the corresponding receivers have $N_1$
and $N_2$ antennas, respectively, the exact characterization of
the DoF region is available for all values of the 4-tuple
$(M_1,N_1,M_2,N_2)$, except if the inequality $\min (M_1, N_1) > N_2
> M_2$ (or its symmetric counterpart obtained by reversing the
user-ordering) holds. This basic result on the two-user IC is then
generalized to the case of the $K$-user IC. More specifically, the DoF
region of the $K$-user MIMO IC is derived for two classes (see Theorem \ref{thm:
K-user IC}), namely, when
(a) all transmitters have equal number of antennas and so do all receivers
 and (b) each transmitter has no fewer antennas than its paired receiver.
These results imply among other things, that over
the $K$-user SISO IC, one can achieve only $1$ sum-DoF when there is
no CSIT, a conclusion that is in sharp contrast with the result that
proves, via interference alignment, the achievability of
$\frac{K}{2}$ sum-DoF under perfect CSIT \cite{Cadambe}.
Taken together, these results provide a strong motivation for
studying the $K$-user interference channel under the realistic
assumption of partial CSIT made available through low-rate feedback
broadcast links from each receiver to all other nodes (cf. \cite{Rajesh_Varanasi_IA_2009}).
Our result on the MIMO IC
is then extended to the $K$-user MIMO X channel in which every
transmitter has a message for every receiver. It is shown that without CSIT,
the $K$-user IC and the $K$-user X channel have identical
DoF regions for the class of networks in which all transmitters have equal number of antennas
and so do all receivers (see Theorem \ref{thm: dof X channel}). Our results on the no-CSIT DoF
regions strictly include all cases of $K$-user MIMO ICs and X channels for which perfect CSIT DoFs
are known from \cite{GouJafar,Jafar_Cadambe_wireless_X_network}.

The first version of this work \cite{Vaze_Dof_initial} and two others \cite{D.Guo, Chiachi2}
obtain results on the DoF regions of the 2-user MIMO IC simultaneously and independently.
In particular, \cite{Chiachi2} considers the i.i.d. Rayleigh fading model which is included
in the models of \cite{Vaze_Dof_initial} and \cite{D.Guo} deals with isotropic fading which is
also included in the fading distribution models of this paper.
Interestingly, \cite{D.Guo, Chiachi2} also provide inner and outer bounds which coincide
with the bounds derived in \cite{Vaze_Dof_initial} (and in this paper).
Hence, \cite{D.Guo, Chiachi2} also provide the exact characterization
of the DoF region of the 2-user MIMO IC, except if $\min(N_1, M_1) > N_2 > M_2$ or its
symmetric counterpart holds but consider fading distributions and additive noise
models that are somewhat less general than the models considered in this work.

In this paper we also study interference networks with cognition.
In particular, the 2-user MIMO CRC, which is an IC in which one
of the transmitters (the cognitive transmitter) is assumed to know the
message of the other transmitter (the primary transmitter) non-causally
\cite{Devroye}. The DoF region of the CRC with perfect CSIT is known
in the literature \cite{Chiachi-Jafar}. In
\cite{Vaze3}, the authors found an achievable sum-DoF for the MIMO CRC
without CSIT. Here, inner and outer bounds on the DoF
region are obtained (see Theorems \ref{thm: inner bound CRC} and
\ref{thm: outer-bound CRC}). These bounds are seen to coincide for
a vast majority of the values of the 4-tuple $(M_1,N_1,M_2,N_2)$, except
when the inequality $\min(N_1,M_1+M_2) > N_2 > M_2$ holds.
The DoF regions for models of cognition where one or more terminals are cognitive
are dealt with for perfect CSIT in \cite{Chiachi-Jafar} and without CSIT by the authors
 in \cite{Vaze_Dof_Cognitive_IC_ISIT}.

The $K$-user MIMO CRC is also studied where there is one primary
transmitter/receiver pair and all other transmitter/receiver pairs are secondary.
The secondary transmitters are assumed to know the message of the primary transmitter non-causally.
The DoF regions of the $K$-user CRC are derived for the two classes wherein (a) all secondary transmitters have no fewer antennas than their paired receivers and (b) all transmitters have $M$ antennas and all receivers have $N$ antennas with $N>M$ (see Theorem \ref{thm: K-user CRC}).

Our results on general $K$-user MIMO networks are enabled by our analysis of the $K$-user MIMO BC.
In addition to showing that that analysis enables us to find the DoF region results of the $K$-user IC, X and CRC networks we also show that it can be used to obtain the DoF region of the $K$-user multi-hop IC in which the $K$ transmit/receive terminals are separated by multiple orthogonal layers of relays, with $K$ relays per layer (see Appendix \ref{app: two_hop_IC}) and where the relays in the last layer are unaware of their outgoing channel matrices.

As another important application of the DoF region characterization of the $K$-user MIMO BC of this paper, the reader is referred to the recent works on the DoF of the $K$-user MISO BC \cite{maddah_ali_tse_delayed_CSIT} and DoF region of the 2-user MIMO BC \cite{Vaze-Varanasi-delay-MIMOBC} with {\em delayed} CSIT where it is shown that delayed CSIT even in i.i.d. fast fading channels can help bridge a significant part of the vast gap between the two extreme cases of perfect CSIT and complete lack of CSIT.

\section{The $K$-user MIMO BC}
\subsection{Channel Model} \label{subsec: BC channel model}

\begin{figure} \centering
\includegraphics[height=3in,width=4in]{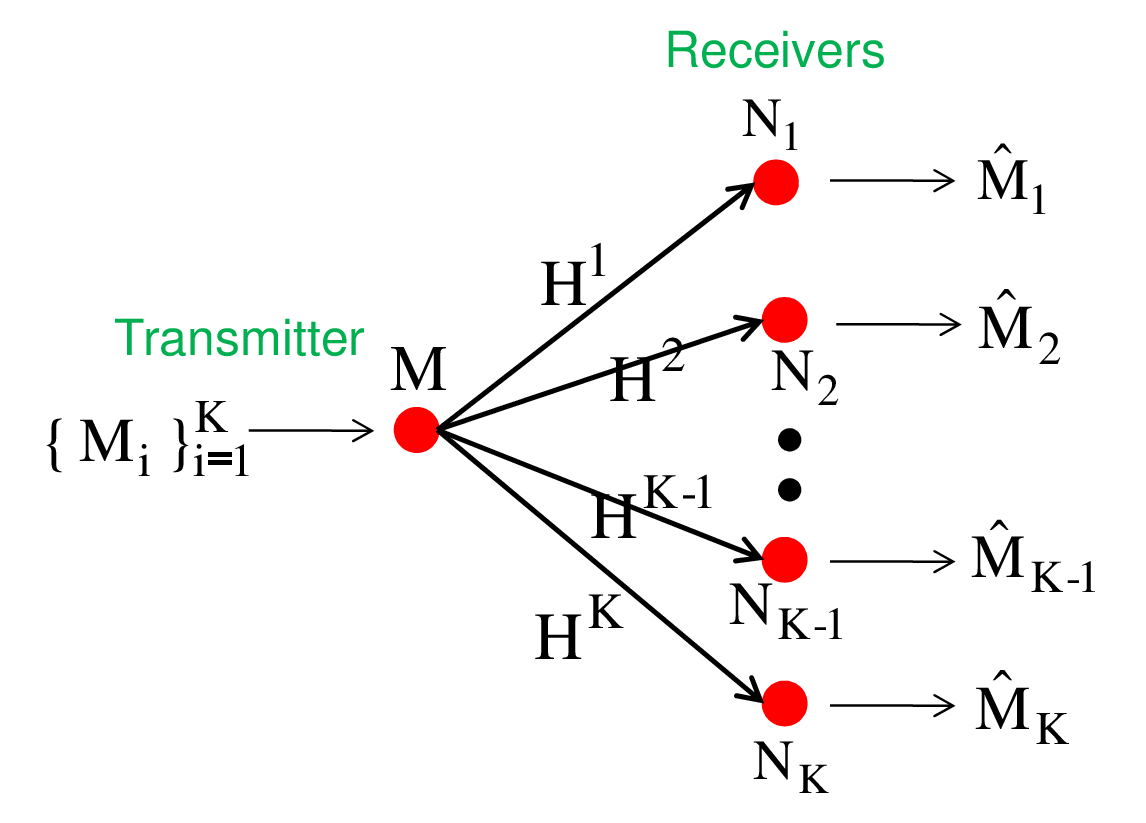}
\caption{The $K$-User MIMO BC.} \label{fig: MIMO_Kuser_BC}
\end{figure}

Consider the $K$-user MIMO BC depicted in Fig. \ref{fig: MIMO_Kuser_BC}
with $M>1$ transmit antennas and users $1$ through $K$ having $N_1$, $\cdots$, $N_K$
receive antennas, respectively. The input-output
relationship at the $i^{th}$ receiver in the $t^{th}$ time slot is given by
\begin{equation}
Y^i(t) = H^i(t) X(t) + W^i(t), \label{eq: BC channel model}
\end{equation}
where, at time $t$, $Y^i(t) \in \mathbb{C}^{N_i \times 1}$
is the signal received by the $i^{th}$ user, $H^i(t) \in
\mathbb{C}^{N_i \times M}$ is the $i^{th}$ user's channel matrix,
$X(t) \in \mathbb{C}^{M \times 1}$ is the signal transmitted under
the power constraint of $\lim_{n \to \infty} \frac{1}{n}
\sum_{t=1}^n \mathbb{E} ||X(t)||^2 \leq P$, and $W^i(t)$ is the
additive noise. All the receivers are assumed to have perfect knowledge
of all the channel matrices instantaneously, an assumption which we will refer to as CSIR; however, the transmitter does not have knowledge of channel realizations but knows the distribution of the channel matrices. Taken together, we refer to these two assumptions simply as the ``no CSIT" assumption. Define ${\rm SNR} = P$.

We now specify the distribution of the additive noise and the
channel matrices for which we state the following two
definitions.

\begin{definition}[Additive white Gaussian noise (AWGN)]
The additive noise is said to be AWGN if $W^i(t)$ are independent and identically distributed (i.i.d.) across time according the zero-mean complex Gaussian distribution with identity covariance matrix denoted as $\mathcal{C}\mathcal{N} \big( 0,I_{N_i} \big)$.
\end{definition}

\begin{definition}[Class of channel distributions $\mathcal{D}_0(M,
\bar{N})$] \label{defn:do} We say that the matrices
$\{H^i(t)\}_{i=1}^K$ follow a distribution of type $\mathcal{D}_0(M,
\bar{N})$ with $\bar{N} = (N_1,N_2,\cdots,N_K)$, if $H^i(t)$ are
i.i.d. across $t$ according to $H^i$ $\forall$ $i$, $\{H^i\}$'s are
independent across users, and $\{H^i\}$'s have i.i.d. rows with
independent channel norms and directions across receiver antennas so
that their distributions can be described as follows. Let $f \in
\mathbb{C}^{1 \times M}$ be a complex-valued unit-norm random row
vector. Consider matrices $\{F^i \in \mathbb{C}^{N_i \times M}
\}_{i=1}^K$ whose rows are all i.i.d. according to $f$. Consider
square diagonal matrices $\{\Lambda^i \in \mathbb{C}^{N_i \times
N_i} \}_{i=1}^K$, where $\Lambda^i$ contains entries
$\{h^{ij}\}_{j=1}^{N_i}$ along its diagonal. The diagonal elements
$\{h^{ij}\}_{i,j}$ are all independent non-negative random
variables, which are also independent of $\{F^i\}$'s. Define random
matrices $H^i = \Lambda^i F^i$ $\forall$ $i$, and assume that for
each $i$, $\{H^i\}$ is full rank with probability $1$ and that each
row has differential entropy greater than $-\infty$.
\end{definition}

The i.i.d. Rayleigh fading wherein all
entries of all channel matrices are i.i.d. standard complex Gaussian
$\mathcal{C}\mathcal{N} \big( 0,1 \big)$ random variables clearly
falls in the above category of distributions
\footnote{If a vector consists of
i.i.d. standard complex Gaussian $\mathcal{C}\mathcal{N} \big( 0,1
\big)$ random variables, then its direction and norm are
independent. Moreover, its differential entropy is
clearly $> -\infty$ and a matrix that consists of such i.i.d.
vectors is full rank with probability $1$.}.

The DoF region of the MIMO BC with AWGN when the channel
matrices follow a distribution of type $\mathcal{D}_0(M, \bar{N})$ is first derived.
This result is then generalized in Section \ref{subsec: BC
generalizations} to prove that the same DoF region applies to
a much wider class of MIMO BCs.


Consider any coding scheme that achieves the rate tuple
$(R_1,R_2,\cdots,R_K)$. Let $M_i$ be the message to be sent to user
$i$ over the blocklength of $n$. We assume that the messages are
independent and message $M_i$ is distributed uniformly over a set of
cardinality $2^{nR_i}$. We say that the rate tuple
$(R_1,R_2,\cdots,R_K)$ is achievable if, at every user, the
probability of error in decoding the respective message goes to zero
as the blocklength $n \to \infty$. Further, since there is no CSIT,
the transmit signal is independent of the actual realizations of the
channel matrices. We define the capacity region $\mathcal{C}(P)$ to
be the set of all achievable rate tuples $(R_1,R_2,\cdots,R_K)$ when
the transmit-power constraint is $P$.

\begin{definition}
\label{defn:dof}
The degrees of freedom (DoF) region represents the set of all $K$ tuples of high ${\rm SNR}$ slopes corresponding to the (achievable) rate-tuples in the capacity region relative to $\log(P)$. It is therefore defined as
follows:
\[ \mathbf{D} = \left\{(d_1, \cdots, d_n) \left| ~ d_i \geq 0 \mbox{
and } \exists ~ (R_1(P), \cdots, R_n(P)) \in \mathcal{C}(P) \; \mbox{
s.t.}\; d_i = \mathrm{MG}(R_i(P)) \hspace{2pt} ~ \forall ~ i
\right. \right\}, \] where the function multiplexing gain $\rm{MG}
(\cdot)$ is defined as $\rm{MG}(x) = \lim_{P \to \infty}
\frac{x}{\log P}$.
\end{definition}
Since a single-antenna point-to-point channel has 1 DoF, the DoF region can be thought of as denoting the set of all highest, simultaneously accessible fractions of spatial signaling dimensions (per channel use) by the users.

It is well-known that the DoF region under the idealized assumption of perfect CSIT for the $K$-user MIMO BC with $M$ transmit antennas and $N_i$ receiver antennas at receiver $i$ is characterized by the single-user bounds $d_i \leq \min\{M,N_i\}$ and the sum-DoF inequality $ \sum_{i=1}^K d_i \leq \min \{ M, \sum_{i=1}^K N_i \}$ (cf. \cite{Caire,Shamai-W-S}).

\subsection{The DoF Region}
We next state our main result about the DoF region of the MIMO BC under the no CSIT assumption.
\begin{theorem} \label{thm: dof region BC}

The DoF region of the $K$-user MIMO BC with AWGN, under the no CSIT assumption
and with the channel matrices having a distribution of type $\mathcal{D}_0(M, \bar{N})$,
is given as
\begin{equation}
\mathbf{D} = \left\{ (d_1, \cdots, d_n) \left| d_i \geq 0 ~ \forall
~ i, ~ \frac{d_1}{\min(M,N_1)} + \frac{d_2}{\min(M,N_2)} + ~ \cdots
~ + \frac{d_K}{\min(M,N_K)} \leq 1 \right. \right\}. \label{eq: dof
region BC}
\end{equation}
\end{theorem}

\begin{IEEEproof}
The above DoF region is achievable by the simple time-division scheme,
and hence, is an inner-bound. The fact that it is also an outer-bound is
proved in detail in Section \ref{subsec: BC proof}, thereby establishing it to be the fundamental DoF region of the MIMO BC.
\end{IEEEproof}

\begin{remark}[Applicability to respective CSIR]
Consider that the receivers know only their own channel matrices
(referred to as ``respective CSIR"). The region in (\ref{eq: dof region BC}) is
still an inner-bound since time-division does not require
receivers to know other users' channel matrices. Moreover, since
(\ref{eq: dof region BC}) is an outer bound with CSIR, it is also an
an outer-bound with respective CSIR. Hence (\ref{eq: dof region BC}) is also the DoF region for the MIMO BC
with respective CSIR.
\end{remark}

\begin{remark}[The loss of DoF]
With perfect CSIT, the sum-DoF of $\min(M,\sum_i N_i)$ can be achieved. According to Theorem \ref{thm: dof region BC} however, with no CSIT, the sum-DoF are only $\min(M,\max_i N_i)$. There is hence a loss of DoF due to the lack of CSIT.
For example, if $M = \sum_i N_i$ and $N_i = 1$ $\forall$ $i$, then the sum-DoF collapse from $M$ to $1$.
\end{remark}

\begin{remark}[The case of partial CSIT]
From a practical perspective, it is important to explore the
possibility of achieving higher DoF, which, as per Theorem \ref{thm:
dof region BC}, is feasible only if there is at least partial CSIT.
In particular, it has been proved that if the quality of CSIT improves at a
sufficient rate with the transmit power, any given sum-DoF up to $\min(M,\sum_i N_i)$
can be achieved \cite{Jindal, Ravindran,Vaze_fb_scaling_GBC}.
\end{remark}

\begin{remark}[Applications of Theorem \ref{thm: dof region BC}]
It turns out that Theorem \ref{thm: dof region BC} is useful in a
variety of settings other than the MIMO BC. For instance, in Section
\ref{sec:networks} we use the result to derive the DoF regions of
certain classes of $K$-user MIMO interference, X and cognitive
interference networks. Moreover, in
\cite{maddah_ali_tse_delayed_CSIT}, the authors consider the
$K$-user MISO Gaussian BC (i.e., single antenna receivers) with
delayed CSI in which every terminal including the transmitter is
assumed to have perfect CSI but with some delay, and every receiver
knows its own channel instantaneously. Using Theorem \ref{thm: dof
region BC} an outer-bound to the DoF region was derived in
\cite{maddah_ali_tse_delayed_CSIT} and was shown to be tight when
$M\geq K$. The outer bound of \cite{maddah_ali_tse_delayed_CSIT} was
extended by the authors again using Theorem \ref{thm: dof region BC}
to the MIMO BC and was shown to be tight for the 2-user case
in \cite{Vaze-Varanasi-delay-MIMOBC}. Another example that
illustrates the application of Theorem \ref{thm: dof region BC} is
given in Appendix \ref{app: two_hop_IC}, which gives the DoF region
of a $K$-user multi-hop interference network wherein the
transmitters wish to send $K$ independent messages to their
respective receivers with the help of multiple layers of relays.
\end{remark}

\subsection{Proof of Theorem \ref{thm: dof region BC}} \label{subsec: BC proof}

We assume without loss of generality that $N_1 \geq N_2 \geq \cdots
\geq N_K$. To obtain the outer-bound, we
enhance the capacity region of the original BC by assuming that
receiver $i$ knows messages $M_{i+1}$ through $M_K$ (denoted as
$M_{i+1:K}$). We then apply Fano's inequality to upper-bound
the achievable rates. Before getting into the details however, we first
introduce some notation.

\emph{\underline{Notation:} } For a column vector $V(t)$ we define
$\mathbf{V} \equiv \mathbf{V}_1^n$ to be a vector $[V^T(1) \cdots V^T(n)]^T $
For a matrix $M(t)$, we define $\mathbf{M}
\equiv \mathbf{M}_1^n$ to be a block-diagonal matrix with entries
$M(1)$, $M(2)$, $\cdots$, $M(n)$ along the diagonal in that order. Define $x_i = \min(M,N_i)$.


Now, by Fano's inequality, we get
\[ R_i \leq \frac{1}{n} I \left( M_i;\mathbf{Y^i} |M_{i+1:K},
\mathbf{H} \right) + \epsilon_n ,\] where $\mathbf{H}$ denotes the
collection of random matrices $\mathbf{H^1}$, $\cdots$, $\mathbf{H^K}$; the
scalar $\epsilon_n$ is such that it goes to zero as $n \to \infty$;
and $M_{K+1:K} = \phi$ denotes some deterministic number. Taking the
limit over $n$, we obtain
\begin{equation}
R_i \leq \lim_{n \to \infty} \frac{1}{n} I \left( M_i;\mathbf{Y^i}
|M_{i+1:K}, \mathbf{H} \right). \label{eq: bound on Ri BC}
\end{equation}
The general approach now is to compute the multiplexing gain of both
the sides of the above equation to obtain bounds on $d_i$, and then,
use these bounds to prove the inequality $\sum_i \frac{d_i}{x_i}
\leq 1$. In order to do so, we prove a key lemma (Lemma \ref{lem: main
ineq}, stated below) that relates the multiplexing gains of certain
differential entropy terms. This lemma can be proved if all the rows of
all channel matrices $\{H^i(t)\}_i$ have identical norms at any given
time $t$. Therefore, in Step I of our proof, we create an `enhanced' channel which has the property required for proving Lemma \ref{lem: main
ineq} and whose capacity region contains that of the original BC. Next, in Step II, Lemma \ref{lem: main ineq} is derived, and finally, in Step III, the required
inequality is proved using the lemma.

\emph{\underline{Step I:} } Note that we may write $H^i(t) =
\Lambda^i(t) F^i(t)$, following our assumption about the
distribution of the channel matrices. Let $h_{\max}(t)$ be the
maximum of all the diagonal entries of $\{\Lambda^i(t)\}_{i=1}^K$.
Define $\tilde{Y}^i (t) = h_{\max}(t) \big( \Lambda^i(t) \big)^{-1}
H^i(t) X(t) + W^i(t)$ and then consider $Y^i(t)' =
\frac{1}{h_{\max}(t)} \Lambda^i(t) \tilde{Y}^i(t) = H^i(t) X(t) +
\frac{1}{h_{\max}(t)} \Lambda^i(t) W^i(t)$.

Conditioned on $\mathbf{H}$, $\frac{1}{h_{\max}(t)} \Lambda^i(t)
W^i(t) \sim \mathcal{C} \mathcal{N} \big( 0,D^i(t) \big)$ where
$D^i(t)$ is an $N_i \times N_i$ diagonal matrix whose $j^{th}$
diagonal entry equals a positive number, $\left(
\frac{h^{ij}(t)}{h_{\max}(t)} \right)^2$, that is less than or equal
to $1$. Hence, if we consider a noise vector $W^i(t)'$ that is
independent of all other transmit-receive signals and the noise
vectors and whose distribution conditioned on $\mathbf{H}$ is given
by $W^i(t)' \sim \mathcal{C}\mathcal{N} \big( 0, D^i(t)' \big)$
where $D^i(t)'$ is a square diagonal matrix such that $D^i(t) +
D^i(t)' = I_{N_i}$, then the signal $Y^i(t)' + W^i(t)'$ is
statistically equivalent to $Y^i(t)$. Hence, we have the following
Markov chain
\[M_i \to X(t) \to \tilde{Y}^i(t) \to Y^i(t)' \to Y^i(t), ~ \forall
~ t,  \] when conditioned on $\mathbf{H}$ and $M_{i+1:K}$. The data
processing inequality \cite{CT} then implies that
\[ I \big( M_i;\mathbf{Y^i} |M_{i+1:K}, \mathbf{H} \big) \leq
I \big( M_i;\mathbf{\tilde{Y}^i} |M_{i+1:K}, \mathbf{H} \big).\]

Note that in going from $Y^i(t)$ to $\tilde{Y}^i(t)$, we have
increased the norms of the rows of $H^i(t)$ while maintaining the
noise statistics unaltered, or equivalently, we have reduced the
variance of the additive noise, and the above inequality says that
this can only increase the mutual information. In what follows, this technique is
referred to as `channel enhancement'. Essentially, this step
loosens the upper-bound obtained through Fano's inequality in a manner such that the
DoF result does not change.

We now have the following bound on $d_i$:
\begin{eqnarray}
d_i = \mathrm{MG}(R_i) & \leq & \mathrm{MG} \left\{ \lim_n
\frac{1}{n} I \Big( M_i;\mathbf{\tilde{Y}^i} |M_{i+1:K}, \mathbf{H}
\Big) \right\}
\nonumber \\
& = & \mathrm{MG}\left\{ \lim_n \frac{1}{n} h \big(
\mathbf{\tilde{Y}^i} |M_{i+1:K}, \mathbf{H} \big) \right\} -
\mathrm{MG} \left\{\lim_n \frac{1}{n} h \big( \mathbf{\tilde{Y}^i} |
M_{i:K}, \mathbf{H} \big) \right\}. \label{eq: bounds on di's}
\end{eqnarray}

\emph{\underline{Step II:} } Recall that $x_i = \min(M,N_i)$. We
have the following key lemma.
\begin{lemma} \label{lem: main ineq}
The inequality below holds for each $i>1$:
\[ \frac{1}{x_{i-1}} \mathrm{MG} \left\{\lim_n \frac{1}{n} h \big(
\mathbf{\tilde{Y}^{i-1}} | M_{i:K}, \mathbf{H} \big) \right\} \leq
\frac{1}{x_i} \mathrm{MG} \left\{\lim_n \frac{1}{n} h \big(
\mathbf{\tilde{Y}^i} | M_{i:K}, \mathbf{H} \big) \right\}.
\]
\end{lemma}
Before proving this lemma, we first show how it is useful.

\emph{\underline{Step III:} } Consider the bound on $d_1$
given by (\ref{eq: bounds on di's}). Since the transmitted signal is
determined by the messages, it is easy to see that
\[ \mathrm{MG} \left\{\lim_n \frac{1}{n} h \big( \mathbf{\tilde{Y}^1} | M_{1:K},
\mathbf{H} \big) \right\} = 0, \] which implies that \[ d_1 \leq
\mathrm{MG}\left\{ \lim_n \frac{1}{n} h \big( \mathbf{\tilde{Y}^1}
|M_{2:K}, \mathbf{H} \big) \right\}. \] Using Lemma \ref{lem: main
ineq} for $i=2$, the following upper-bound can be derived
\begin{equation}
\frac{d_1}{x_1} \leq \frac{1}{x_1} \mathrm{MG}\left\{ \lim_n
\frac{1}{n} h \big( \mathbf{\tilde{Y}^1} |M_{2:K}, \mathbf{H} \big)
\right\} \leq \frac{1}{x_2} \mathrm{MG}\left\{ \lim_n \frac{1}{n} h
\big( \mathbf{\tilde{Y}^2} |M_{2:K}, \mathbf{H} \big) \right\}.
\label{eq: bound d1/x1}
\end{equation}
Now, using the bound on $d_2$ obtained from (\ref{eq: bounds on
di's}) and the above inequality, we have
\begin{eqnarray}
\frac{d_2}{x_2} \leq \frac{1}{x_2} \mathrm{MG}\left\{ \lim_n
\frac{1}{n} h \big( \mathbf{\tilde{Y}^2} |M_{3:K}, \mathbf{H} \big)
\right\} - \underbrace{ \frac{1}{x_2} \mathrm{MG} \left\{\lim_n
\frac{1}{n} h \big( \mathbf{ \tilde{Y}^2} | M_{2:K}, \mathbf{H}
\big) \right\} }_{\geq \frac{d_1}{x_1} } \nonumber \\
\Rightarrow \frac{d_1}{x_1} + \frac{d_2}{x_2} \leq  \frac{1}{x_2}
\mathrm{MG}\left\{ \lim_n \frac{1}{n} h \big( \mathbf{\tilde{Y}^2}
|M_{3:K}, \mathbf{H} \big) \right\}. \label{eq: bound on
d1/x1+d2/x2}
\end{eqnarray}
Again, invoke the lemma with $i=3$ to get
\[\frac{d_1}{x_1} + \frac{d_2}{x_2} \leq \frac{1}{x_2}
\mathrm{MG}\left\{ \lim_n \frac{1}{n} h \big( \mathbf{\tilde{Y}^2}
|M_{3:K}, \mathbf{H} \big) \right\} \leq \frac{1}{x_3}
\mathrm{MG}\left\{ \lim_n \frac{1}{n} h \big( \mathbf{\tilde{Y}^3}
|M_{3:K}, \mathbf{H} \big) \right\}. \] We will now use the bound on
$d_3$ given by equation (\ref{eq: bounds on di's}) and the
above inequality to derive
\begin{eqnarray}
\lefteqn{ \hspace{-2cm} \frac{d_3}{x_3} \leq \frac{1}{x_3}
\mathrm{MG}\left\{ \lim_n \frac{1}{n} h \big( \mathbf{\tilde{Y}^3}
|M_{4:K}, \mathbf{H} \big) \right\} - \frac{1}{x_3}
\mathrm{MG}\left\{ \lim_n \frac{1}{n}
h \big( \mathbf{\tilde{Y}^3} |M_{3:K}, \mathbf{H} \big) \right\} }\nonumber \\
&& {} \hspace{-1cm} \Rightarrow \frac{d_1}{x_1} + \frac{d_2}{x_2} +
\frac{d_3}{x_3} \leq \frac{1}{x_3} \mathrm{MG}\left\{ \lim_n
\frac{1}{n} h \big( \mathbf{\tilde{Y}^3} |M_{4:K}, \mathbf{H} \big)
\right\}. \label{eq: bound on d1/x1 + d2/x2 + d3/x3}
\end{eqnarray}
Working successively this way, we finally get
\begin{equation}
\frac{d_1}{x_1} + \frac{d_2}{x_2} + \cdots + \frac{d_K}{x_K} \leq
\frac{1}{x_K} \mathrm{MG}\left\{ \lim_n \frac{1}{n} h \big(
\mathbf{\tilde{Y}^K} |M_{K+1:K}, \mathbf{H} \big) \right\}
\label{eq: bound on d1/x1 + d2/x2 + .... + dK/xK}
\end{equation}
at the last stage. Recall here that $M_{K+1:K} = \phi$, a
deterministic number. Since the MIMO channel with $M$ transmit and
$N$ receive antennas can have at most $\min(M,N)$ DoF, we get
\begin{equation}
\frac{d_1}{x_1} + \frac{d_2}{x_2} + \cdots + \frac{d_K}{x_K} \leq
\frac{1}{x_K} \mathrm{MG}\left\{ \lim_n \frac{1}{n} h \big( \mathbf{
\tilde{Y}^K } |\phi, \mathbf{H} \big) \right\} \leq \frac{1}{x_K}
\min(M,N_K) = 1,
\end{equation}
as required.

It can be noted that the key to the above proof is Lemma \ref{lem:
main ineq} which relates the multiplexing gains of certain
differential entropy terms. It is also important that we weigh these
multiplexing gains by appropriate fractions, namely, $\frac{1}{x_i}$
and $\frac{1}{x_{i-1}}$ before we bound the one by the other. It is
not hard to see that if these fractions are different from the ones
we have here, we may not necessarily get the tightest result.

Our approach detailed above when specialized to the case of $K = 2$
may roughly resemble the one of \cite{Lapidoth}. Recall first however that
\cite{Lapidoth} considers the real-valued Gaussian
partial-CSIT BC with $2$ transmit antennas and $2$ single-antenna
receivers. Some sort of a counterpart of
Lemma \ref{lem: main ineq} is proved therein, where the term $ \mathrm{MG}
\left\{\lim_n \frac{1}{n} h \big( \mathbf{\tilde{Y}^i} | M_{i:K},
\mathbf{H} \big) \right\}$ is lower-bounded by a fraction
$\frac{1}{2}$ times the term $\mathrm{MG} \left\{\lim_n \frac{1}{n}
h \big( \mathbf{ \tilde{Y}^{i-1} } | M_{i:K}, \mathbf{H} \big)
\right\}$ \footnote{Note that it is only for the sake of
illustration that we use the quantities, $\mathrm{MG} \left\{\lim_n
\frac{1}{n} h \big( \mathbf{\tilde{Y}^i} | M_{i:K}, \mathbf{H} \big)
\right\}$ and $\mathrm{MG} \left\{\lim_n \frac{1}{n} h \big(
\mathbf{\tilde{Y}^{i-1}} | M_{i:K}, \mathbf{H} \big) \right\}$, in
our explanation here. The actual terms involved in \cite{Lapidoth}
are different and more complicated than (but in some sense the
direct analogues of) the ones we have here due to the consideration
of the more general partial-CSIT case. To even get this factor of
$\frac{1}{2}$ under partial CSIT requires considerable work \cite{Lapidoth}.}.
This yields them a bound $d_1 + d_2 \leq \frac{2}{3}$, whose
tightness is still unknown. Here, for our no-CSIT model, we are
interested in proving the tightest result, namely, the bound $d_1 +
d_2 \leq \frac{1}{2}$ for real-valued channels or, equivalently,
$d_1+ d_2 \leq 1$ for complex-valued channels. This requires
us to get the above fraction to be unity (since $x_1 = x_2 = 1$ in
this particular case), which is what Lemma \ref{lem: main
ineq} provides. It remains only to prove this lemma.

\emph{\underline{Step II -- Proof of Lemma \ref{lem: main ineq}:} }
Let us first introduce some notation.

\emph{\underline{Notations:} } Let $V_i(t)$ be the $i^{th}$ element
of the column vector $V(t)$. We define $(\mathbf{V}_i)_1^n \equiv
\mathbf{V}_i$ to be the vector $ [ V_i(1), V_i(2), \cdots , V_i(n) ]^T$. For example,
if we consider the vector $\tilde{Y}^i(t)$, then its $j^{th}$ entry
is denoted by $\tilde{Y}^i_j(t)$, and the vector
$\mathbf{\tilde{Y}^i}_j $ is defined as $ [\tilde{Y}^i_j(1), \tilde{Y}^i_j(2), \cdots, \tilde{Y}^i_j(n)]^T$.

The proof consists of two steps: initially, it is proved that, without
loss of generality, we may assume $N_{i-1}$, $N_i \leq M$; later, we
will prove the result for the case of $N_{i-1}, ~ N_i \leq M$.

\emph{\underline{Step II.a:} } Suppose $N_i > M$. Then we show
that
\begin{equation}
\mathrm{MG} \left\{\lim_n \frac{1}{n} h \big( \mathbf{\tilde{Y}^i} |
M_{i:K}, \mathbf{H} \big) \right\} = \mathrm{MG} \left\{\lim_n
\frac{1}{n} h \big( \mathbf{\tilde{Y}^i}_1,  \mathbf{\tilde{Y}^i}_2,
\cdots, \mathbf{\tilde{Y}^i}_M | M_{i:K}, \mathbf{H} \big) \right\}.
\label{eq: Step II.a equality}
\end{equation}
To this end, consider $\tilde{Y}^i(t) = h_{\max}(t) F^i(t) X(t) +
W^i(t)$. Since, by assumption, the matrix $F^i(t)$ is full
rank with probability $1$ its first $M$ rows can be taken
to be linearly independent. Let the $M \times M$ matrix formed out
of the first $M$ rows of $F^i(t)$ be $F^i_{1:M}(t)$. Then given the
first $M$ entries $\tilde{Y}^i_{1:M}(t)$ of $\tilde{Y}^i(t)$, a
noisy version of the transmit signal, namely, $X'(t) = \big(
F^i_{1:M}(t) \big)^{-1} \tilde{Y}^i_{1:M}(t) = X(t) + \big(
F^i_{1:M}(t) \big)^{-1} W^i_{1:M}(t)$ can be computed. Therefore, we
have
\begin{eqnarray*}
\lefteqn{ \mathrm{MG} \left\{ h \big(\tilde{Y}^i(t) | \mathbf{H},
M_{i:K} \big) \right\} } \\
& \hspace{-1.7cm} = & \hspace{-1cm} \mathrm{MG} \left\{ h \big(
\tilde{Y}^i_{1:M}(t) \big| \mathbf{H}, M_{i:K} \big) \right\} +
\mathrm{MG} \left\{ h \big( \tilde{Y}^i_{M+1:N_i}(t) \big|
\tilde{Y}^i_{1:M}(t), \mathbf{H}, M_{i:K} \big) \right\} \\
& \hspace{-1.7cm} = & \hspace{-1cm} \mathrm{MG} \left\{ h \big(
\tilde{Y}^i_{1:M}(t) \big| \mathbf{H}, M_{i:K} \big) \right\} +
\mathrm{MG} \left\{ h \big( \tilde{Y}^i_{M+1:N_i}(t) - h_{\max}(t)
F^i_{M+1:N_i}(t) X'(t) \big| \tilde{Y}^i_{1:M}(t), \mathbf{H}, M_{i:K} \big) \right\} \\
& \hspace{-1.7cm} = & \hspace{-1cm} \mathrm{MG} \left\{ h \big(
\tilde{Y}^i_{1:M}(t) \big| \mathbf{H}, M_{i:K} \big) \right\} +
\mathrm{MG} \left\{ h \big( \tilde{W}^i_{M+1:N_i}(t) - h_{\max}(t)
F^i_{M+1:N_i}(t) \big( F^i_{1:M}(t) \big)^{-1} W^i_{1:M}(t) \big| \mathbf{H}, M_{i:K} \big) \right\} \\
& \hspace{-1.7cm} = & \hspace{-1cm} \mathrm{MG} \left\{ h \big(
\tilde{Y}^i_{1:M}(t) \big| \mathbf{H}, M_{i:K} \big) \right\}.
\end{eqnarray*}
The equality in (\ref{eq: Step II.a equality}) now follows.
Similarly, we can handle the case of $N_{i-1} > M$.
Hence, we may assume that $N_i, ~ N_{i-1} \leq
M$, which of course implies $x_i = N_i$ and $x_{i-1} = N_{i-1} $.

\emph{\underline{Step II.b:} } We will prove that
\begin{equation}
\label{eq:eqtolemma1}
 \frac{1}{N_{i}}
\cdot h(\mathbf{\tilde{Y}^i} | M_{i:K}, \mathbf{H}) \geq
\frac{1}{N_{i-1}} \cdot  h(\mathbf{\tilde{Y}^{i-1}} | M_{i:K},
\mathbf{H})
\end{equation}
from which the proof of Lemma 1 follows.

Consider the two sets of random variables $\{
\mathbf{\tilde{Y}^{i-1}}_1,~ \mathbf{\tilde{Y}^{i-1}}_2, ~ \cdots, ~
\mathbf{\tilde{Y}^{i-1}}_{N_{i-1}}\}$ and
$\{\mathbf{\tilde{Y}^i}_1,~ \mathbf{\tilde{Y}^i}_2, ~\cdots, ~
\mathbf{\tilde{Y}^i}_{N_i}\}$. For a given integer $m$ such that $0
< m \leq \min(N_i,N_{i-1})$, it follows by symmetry that the joint
distribution of any $m$ random variables chosen from the first set of $N_{i-1}$ random variables
is identical to that of any $m$ random variables chosen from the second set of $N_i$
random variables. In fact, this is true of their conditional joint
distributions as well, if we condition them on the same set of
random variables. We refer to this property
as the `statistical equivalence of the involved random variables'. Using this
property, we get
\begin{eqnarray}
h(\mathbf{\tilde{Y}^i}|M_{i:K}, \mathbf{H}) & = & \sum_{j=1}^{N_i}
\left\{ h(\mathbf{\tilde{Y}^i}_j, \mathbf{\tilde{Y}^i}_{j-1} \cdots,
\mathbf{\tilde{Y}^i}_1|M_{i:K}, \mathbf{H}) -
h(\mathbf{\tilde{Y}^i}_{j-1}, \mathbf{\tilde{Y}^i}_{j-2}
\cdots, \mathbf{\tilde{Y}^i}_1|M_{i:K}, \mathbf{H}) \right\} \nonumber \\
& = & \sum_{j=1}^{N_i} \left\{ h(\mathbf{\tilde{Y}^i}_{N_i},
\mathbf{\tilde{Y}^i}_{j-1} \cdots, \mathbf{\tilde{Y}^i}_1|M_{i:K},
\mathbf{H}) - h(\mathbf{\tilde{Y}^i}_{j-1},
\mathbf{\tilde{Y}^i}_{j-2} \cdots, \mathbf{\tilde{Y}^i}_1|M_{i:K},
\mathbf{H}) \right\} \label{eq:Step II.b Z manipu 1} \\
& = & \sum_{j=1}^{N_i} \left\{ h(\mathbf{\tilde{Y}^i}_{N_i}|
\mathbf{\tilde{Y}^i}_{j-1} \cdots, \mathbf{\tilde{Y}^i}_1, M_{i:K},
\mathbf{H}) \right\} \nonumber \\
& \geq & \sum_{j=1}^{N_i} \left\{ h(\mathbf{\tilde{Y}^i}_{N_i}|
\mathbf{\tilde{Y}^i}_{N_i-1} \cdots, \mathbf{\tilde{Y}^i}_1,
M_{i:K},\mathbf{H}) \right\} \label{eq:Step II.b Z manipu 2} \\
& = & N_i \cdot h(\mathbf{\tilde{Y}^i}_{N_i}|
\mathbf{\tilde{Y}^i}_{N_i-1} \cdots, \mathbf{\tilde{Y}^i}_1,
M_{i:K}, \mathbf{H}), \label{eq:Step II.b Z manipu 3}
\end{eqnarray}
where the equality (\ref{eq:Step II.b Z manipu 1}) follows by the
property of the statistical equivalence of the involved random variables
and inequality (\ref{eq:Step
II.b Z manipu 2}) follows since conditioning reduces entropy. Again,
the application of these two ideas and inequality (\ref{eq:Step II.b
Z manipu 3}) gives us
\begin{eqnarray}
\lefteqn{ \hspace{-2cm} (N_{i-1} - N_i) \cdot
h(\mathbf{\tilde{Y}^i}|M_{i:K}, \mathbf{H}) \geq  (N_{i-1} - N_i)
\cdot N_i \cdot h(\mathbf{\tilde{Y}^i}_{N_i} | M_{i:K}, \mathbf{H},
\mathbf{\tilde{Y}^i}_{1}, \cdots, \mathbf{\tilde{Y}^i}_{N_i-1}) } \nonumber \\
& = & (N_{i-1} - N_i) \cdot N_i \cdot
h(\mathbf{\tilde{Y}^{i-1}}_{N_i+1} | M_{i:K}, \mathbf{H},
\mathbf{\tilde{Y}^{i-1}}_2, \mathbf{\tilde{Y}^{i-1}}_3,
\cdots, \mathbf{\tilde{Y}^{i-1}}_{N_i}) \nonumber \\
& \geq & N_i \cdot (N_{i-1} - N_i) \cdot
h(\mathbf{\tilde{Y}^{i-1}}_{N_i+1} | M_{i:K}, \mathbf{H},
\mathbf{\tilde{Y}^{i-1}}_1, \mathbf{\tilde{Y}^{i-1}}_2, \cdots,
\mathbf{\tilde{Y}^{i-1}}_{N_i}) \nonumber \\
& \geq & N_i \cdot h(\mathbf{\tilde{Y}^{i-1}}_{N_i+1},
\mathbf{\tilde{Y}^{i-1}}_{N_i+2}, \cdots,
\mathbf{\tilde{Y}^{i-1}}_{N_{i-1}} | M_{i:K},
\mathbf{H},\mathbf{\tilde{Y}^{i-1}}_1, \mathbf{\tilde{Y}^{i-1}}_2,
\cdots, \mathbf{\tilde{Y}^{i-1}}_{N_i}) \label{eq: Step II.b Z to Y}
\end{eqnarray}
Also, we have
\begin{equation}
h(\mathbf{\tilde{Y}^i}|M_{i:K},\mathbf{H}) =
h(\mathbf{\tilde{Y}^{i-1}}_1, \mathbf{\tilde{Y}^{i-1}}_2, \cdots,
\mathbf{\tilde{Y}^{i-1}}_{N_i} |M_{i:K}, \mathbf{H}).
\label{eq:_joint entropy equality}
\end{equation}
Adding $N_i$ times equation (\ref{eq:_joint entropy equality}) to
inequality (\ref{eq: Step II.b Z to Y}) yields us the required
result of (\ref{eq:eqtolemma1}).  \IEEEQED

\begin{remark}
The class of MIMO BCs considered here does not fall into any
of the special categories, such as, degraded, less noisy, or more
capable BCs (cf. \cite{Gamal_class_of_BC} and the references therein) whose capacity regions are
known. Nevertheless, the above theorem gives us the rate at which the capacity
region scales with SNR for this class of BCs\footnote{The non-degraded nature of the present model
arises due to the magnitude scaling factors, one associated with each
receive antenna. In Step I of the proof which creates an enhanced BC, these are absorbed by the
receivers and then the additive noise is reduced to create a degraded channel.}.
\end{remark}

\begin{remark}
\label{rem:alternative}
The result of Theorem \ref{thm: dof region BC} in the special case of $K = 2$ and
i.i.d. Rayleigh fading model was obtained in \cite{Chiachi}. As pointed out
in the review of this paper, the channel enhancement step (Step I) of the proof of
Theorem \ref{thm: dof region BC} to create a degraded BC whose capacity region is an outer bound
to the BC of interest and an easy extension of the proof of \cite{Chiachi} to accommodate the degraded
model where each receive antenna sees an i.i.d. channel vector can together be used to provide an
alternative proof of the 2-user DoF region for the class of BCs of
Theorem \ref{thm: dof region BC}. Note however, that the proof provided here is stronger
(even considering just the 2-user case), and consequently can, for example, be extended to the case of isotropic
fading which is included in a model for which the DoF result is given in Theorem \ref{thm:D2} (see Remark \ref{rem:comparison} for further details).
\end{remark}

Three simple variants of Theorem \ref{thm: dof region BC} are given in the following remarks.

\begin{remark}[Channel norm information at the transmitter]
Consider the MIMO BCs of Theorem \ref{thm: dof region BC}. Suppose however that the transmitter has perfect knowledge of the norms of the all rows of all channel matrices, i.e., at time $t$, the transmitter knows $\{h^{ij}(t)\}$ $\forall$ $i,j,$ and $t$. However, it has no knowledge about the instantaneous realizations of matrices $\{F^i(t)\}$ $\forall i$ and $t$ (as before, it just knows their distribution). With the aim of determining the DoF region under this setting (where the transmit signal $X(t)$ can depend on the channel norms), consider the following argument: Step I remains valid and the analysis beyond this step depends only on the sequence $\{h_{\max}(t)\}_t$. Steps II.a is insensitive to CSIT and Step III is valid as long as Step II.b is. Now Step II.b follows because of the property of statistical equivalence of involved random variables even though $X(t)$ is dependent on the sequence $\{h_{\max}(t)\}_t$. This argument implies that the DoF region remains unaltered even though the transmitter has channel norm information. This  result underscores the importance of channel direction information.

\end{remark}

\begin{remark}[Channel matrices with i.i.d. rows and additive noise is non-Gaussian]\

Consider the subclass $\mathcal{D}_{-1} (M,\bar{N}) \subset \mathcal{D}_0 (M,\bar{N}) $
of fading matrix distributions in which all rows of all channel
matrices $\{H^i(t)\}$ are i.i.d. with differential entropy greater than $-\infty$ and
the channel matrices are full rank with probability $1$. However, while the additive noise random variables
are assumed to be i.i.d. across receive antennas and across time, they may have any arbitrary distribution with zero mean, unit variance, and differential entropy greater than $-\infty$ (this type of additive noise is referred to as AWN). Under these assumptions, Step I in the proof of Theorem \ref{thm: dof region BC} is
not needed and moreover, it is the only step that depends on the assumption of additive noise being Gaussian.
Consequently, the DoF region without CSIT of the MIMO BC with AWN, but with fading matrices distributed according to $\mathcal{D}_{-1} (M,\bar{N}) $, is the same as the DoF region given in equation (\ref{eq: dof
region BC}) of Theorem \ref{thm: dof region BC}.
\end{remark}

\begin{remark}[Channel matrices correlated across time] \label{remark:correlated}
Let the entries of $H^i(1)$ be i.i.d. $\sim \mathcal{C} \mathcal{N}
(0,1)$ $\forall$ $i$. Let $[H^i(t)]_{jk}$ be the
$(j,k)^{\mathrm{th}}$ element of $H^i(t)$. For all $t > 1$, the
distribution of $H^i(t)$ is defined in the following manner:
$[H^i(t)]_{jk} = \rho \cdot [H^i(t-1)]_{jk} + \sigma \cdot
n^i_{jk}(t)$ where $n^i_{jk}(t) \sim \mathcal{C} \mathcal{N} (0,1)$
i.i.d. across $i$, $j$, $k$, and $t$ and are also independent of
$[H^i(t-1)]_{jk}$; and $\rho, \sigma \in \mathbb{R}$, where $\rho^2
+ \sigma^2 = 1$ and $|\rho| \not= 1$. Let the noise be AWGN.
For this class of MIMO BCs Step I is again not needed and the property
of statistical equivalence of the involved random variables, which is critical
for Step II.b to follow, holds by assumption of distribution of fading matrices.
Hence, the DoF region in this case is also equal to the region defined
in equation (\ref{eq: dof region BC}) of Theorem \ref{thm: dof region BC}.
\end{remark}

Hence, even if the channel matrices are correlated
across time, the DoF of the channel remain
unchanged. It is interesting to contrast this statement with \cite{Jafar_Gou_blind_IA_2010,
Jafar_correlations} where it is shown that if the channel matrices are correlated across time
in some specific manner, then it is possible to (strictly) enhance the DoF
region. Thus, the results of \cite{Jafar_Gou_blind_IA_2010, Jafar_correlations}
may seem to contradict Remark \ref{remark:correlated}. This apparent contradiction
is easily resolved by observing that the staggered block-fading model of
\cite{Jafar_Gou_blind_IA_2010, Jafar_correlations} in which the channel matrices remain constant over the coherence blocks
of length $>$ 1 (and moreover, the boundaries of the coherence blocks of different channel
matrices are suitably misaligned) -- which in turn is critical to
achieving the strictly bigger DoF region -- does not belong to the class of fading model considered in Remark
\ref{remark:correlated} in which the channel strictly varies at each time
instant\footnote{The case of $|\rho| = 1$ is not addressed by Remark \ref{remark:correlated} because
in this case the channel is not fast fading.}.


\subsection{Generalizations of Theorem \ref{thm: dof region BC}} \label{subsec: BC generalizations}
In this section we extend Theorem \ref{thm: dof region BC}
to include a wider class of distributions of the channel matrices and the additive
noises. Towards this end, note that up to Step II.a, we only need the channel matrices to be
invertible with probability $1$, while Step III follows as long as
Lemma \ref{lem: main ineq} holds. Now Step II.b rests on the statistical
equivalence of the involved random variables and hence
depends critically on the distributions of the channel matrices and
the additive noises. Through a detailed look at this step of the proof,
Theorem \ref{thm: dof region BC} can be generalized as discussed below.
Before stating the main results of this section, we define the first
more general class of fading distributions.

\begin{definition}[Fading distributions $\mathcal{D}_1 (M,\bar{N})$ with dependent channel norms and
directions] \label{def: D1}

Let $H^i$ be a representative element for $H^i(t)$, i.e.,
$\{H^i(t)\}_t$ are taken to be i.i.d. (across $t$) according to
$H^i$. As before, let $H^i = \Lambda^i F^i$. The rows of
$\{F^i\}_{i=1}^K$ are random vectors with the following property:
for any $m>0$, if we pick $m$ (distinct) rows (not necessarily from
the same matrix) out of the total $\sum_i N_i$ random row vectors,
then the joint distribution of these does not depend on which
particular $m$ row vectors have been chosen. Further, consider the
distribution of $h_{\max} = \max_{i,j} h^{ij}$, conditioned on $m$
(distinct) row vectors picked from the total $\sum_i N_i$ rows of
$\{F^i\}_{i=1}^K$. It is assumed that this conditional joint
distribution does not depend on which particular $m$ rows that have
been picked. The channel matrices are assumed to be full rank and
each row of these has differential entropy greater than $-\infty$.
If $H^i(t) \sim $ i.i.d. (across $t$) $H^i$ 
then the channel matrices 
are said to follow a distribution of type $\mathcal{D}_1 (M,\bar{N})$. Note that $\mathcal{D}_1 (M,\bar{N}) \supset\mathcal{D}_0 (M,\bar{N})$.
\end{definition}

We also need the following definition.
\begin{definition}[Additive colored Gaussian noise (ACGN)]
The noise is said to be ACGN if $W^i(t)$ are i.i.d. (across time)
according to a distribution $\mathcal{C}\mathcal{N}(0,\Sigma_i)$,
where $\Sigma_i$ is $N_i \times N_i$ positive definite matrix.
\end{definition}
Clearly, AWGN is a special instance of ACGN. The following theorem on the MIMO BC with ACGN and with fading distributions in $\mathcal{D}_1(M,\bar{N})$ is more general than Theorem \ref{thm: dof region BC}.


\begin{theorem} \label{thm:D1}
If the channel matrices follow a distribution of type $\mathcal{D}_1(M,\bar{N})$,
then the DoF region of the $K$-user no-CSIT MIMO BC with ACGN is equal to the region defined in equation (\ref{eq: dof region BC}) of Theorem \ref{thm: dof region BC}.
\end{theorem}
\begin{IEEEproof}
Achievability again follows using time-division. The proof that (\ref{eq: dof region BC}) is also an outer-bound is given in Appendix \ref{app: proof of thm:D1}.
\end{IEEEproof}

Examples of the significance of the above generalization are given in the following three remarks.

\begin{remark}[Rician fading]
As mentioned earlier, the most common assumption about fading distribution is
i.i.d. Rayleigh fading where the elements of the channel matrices
are i.i.d. $\mathcal{C}\mathcal{N}(0,1)$ random variables and the channel
matrices are i.i.d. across time. This type of fading falls under the
category $\mathcal{D}_0(M, \bar{N})$ considered in Section
\ref{subsec: BC channel model}. Consider now the more general
Rician fading model where all the entries of the fading channel matrices are now i.i.d.
$\mathcal{C}\mathcal{N}(\eta,\sigma^2)$ random variables where $|\eta|^2 +
\sigma^2 < \infty$. Evidently, the norm and direction corresponding to any row of the channel matrix,
are {\em not} independent. However,
this distribution falls under the category $\mathcal{D}_1
(M,\bar{N})$ and thus the no CSIT DoF region is known from Theorem \ref{thm:D1}.
\end{remark}

\begin{remark}[Correlated fading]
Theorem \ref{thm:D1} addresses
the case of correlated Rayleigh fading with separable correlations
(cf. \cite{Verdu_power_offset}). Under this type of distribution, we can
write $H^i(t) = (A^i_R) H^i_w(t) (A_T)$ where $A^i_R \in
\mathbb{C}^{N_i \times N_i}$ and $A_T \in \mathbb{C}^{M \times M}$
are fixed invertible square matrices, and the matrix $H^i_w(t)$
contains i.i.d. $\mathcal{C}\mathcal{N}(0,1)$ entries. The receiver
can multiply the received signal by $(A^i_R)^{-1}$ to get rid of
$A^i_R$ and this invertible transformation will not change the
involved mutual informations. However, it will make the additive noise at the
receivers colored (if they were not colored to begin with). Moreover, after this transformation, the effective fading
channel matrices are $H^i_w(t) (A_T)$, and this distribution falls
under the category $\mathcal{D}_1 (M, \bar{N})$. Thus the DoF region of the MIMO BC
with such correlated fading (and ACGN) is given by Theorem \ref{thm:D1}.

\end{remark}

Next we consider a different generalization of Theorem \ref{thm: dof region BC} that subsumes the isotropic fading model.

\begin{definition} \label{def: thm 2.5}
The fading channel distribution is said to be of
type $\mathcal{D}_2 (M,\bar{N})$ if the following assumptions hold:
Let $H^i$ be a representative element for $H^i(t)$, i.e.,
$\{H^i(t)\}_t$ are taken to be i.i.d. (across $t$) according to
$H^i$. Further, let the singular-value decomposition \cite{Horn-Johnson} of $H^i$ be $H^i = U^i \Lambda^i (V^i)^*$ be
the singular-value decomposition of $H^i$ where $U^i$ is $N_i \times
N_i$ unitary matrix, $\Lambda^i$ is $N_i \times \min(M,N_i)$
diagonal\footnote{If $N_i > M$, then we say that $\Lambda^i$ is
diagonal if the square matrix formed out of the first $M$ rows of
$\Lambda^i$ is diagonal and last $N_i-M$ rows have all zero entries.}
matrix containing ordered singular values, and $V^i$ is $M
\times \min(M,N_i)$ semi-unitary matrix (i.e., $(V^i)^* V^i = I$).
Now, $\{V^i\}$'s can have any arbitrary joint distribution with the
following property: For any $i$, $j$, and $m$ with $i \not= j$ and
$m \leq \min(N_i,N_j)$, any set of $m$ (distinct) columns picked
from $V^i$ has the same joint distribution as the set of $m$
(distinct) columns picked from $V^j$. Further, we let all
$\{\Lambda^i\}$'s to be independent of $\{V^i\}$'s with some joint
distribution. $\{U^i\}$'s can be arbitrarily dependent on
$\{\Lambda^i\}$'s and $\{V^i\}$'s. Also, $H^i(t) \sim $ i.i.d. (across
$t$) according to the distribution of $H^i$.
\end{definition}

\begin{remark}[Isotropic fading]
The class of fading distributions $\mathcal{D}_2 (M,\bar{N})$ includes
the isotropic fading model considered in \cite{D.Guo} since by definition
in this model the  singular-value decomposition  $H^i = U^i \Lambda^i (V^i)^*$ yields
$V^i$ to be an isotropically distributed semi-unitary matrix and matrices $U^i$ and
$\Lambda^i$ are independent of $V^i$. The next theorem shows that the no CSIT DoF region of the MIMO BC with AWGN and with fading distribution in  $\mathcal{D}_2 (M,\bar{N})$ is the same as that of Theorem \ref{thm: dof region BC}.
\end{remark}

\begin{theorem}
\label{thm:D2}
The no-CSIT DoF region of the $K$-user MIMO BC with AWGN when the fading channel matrices are distributed as in $\mathcal{D}_2 (M,\bar{N})$ is the same as the DoF region defined by equation (\ref{eq: dof region BC}) of Theorem \ref{thm: dof region BC}.
\end{theorem}
\begin{IEEEproof}
The receiver can multiply the received signal by the unitary matrix
$(U^i)^*$ and this unitary transformation will not change the
involved mutual informations. If $N_i>M$ then the last $N_i-M$
antennas (after the above unitary transformation) would receive only
noise and hence can be ignored. Therefore, after this step, we
may assume that $N_i \leq M$ $\forall$ $i$ and $H^i =
\tilde{\Lambda}^i (V^i)^*$, where $\tilde{\Lambda}^i$ is the square
diagonal matrix formed out of first $M$ rows of $\Lambda^i$. Now the
problem is similar to that of Theorem \ref{thm: dof region BC}.
\end{IEEEproof}

\begin{remark} \label{rem: comparison BC with other results}
The results of \cite{Chiachi, Chiachi2} are applicable for the specific case of two-user Gaussian MIMO BC with AWGN and i.i.d. Rayleigh fading. In contrast, Theorems \ref{thm: dof region BC}-\ref{thm:D2} of this paper apply to the general case of $K \geq 2$ and to a wider class of distributions of channel matrices and additive noises.
The results of this section are also more
general than the result on the DoF region of the MISO BC for
isotropic fading considered in \cite{Jafar-Goldsmith} with
respective CSIR. The extension of the proof technique therein to the
case of multiple-antenna receivers having perfect knowledge of all
channel matrices is not known. Moreover, in \cite{Lapidoth}, the
authors consider the case of partial CSIT in context of a
real-valued BC with 2 transmit antennas and 2 single-antenna
receivers and provide an outer-bound. But the tightness of their
bound is not known in general. In contrast, the results of this
section are valid for the general $K$-user MIMO BC but under the
more restricted no-CSIT assumption.
\end{remark}

\begin{remark} \label{rem:comparison}
Remark \ref{rem:alternative} points to an alternative derivation for the DoF region of the 2-user
MIMO BCs for which the channel matrices follow a distribution of type $\mathcal{D}_0 $. Such
an approach can also be adopted to prove the 2-user version of Theorem \ref{thm:D1} (i.e.,
for channel matrices having distribution of type $\mathcal{D}_1 $). However, it
cannot be extended to BCs in $\mathcal{D}_2$.
This is because the proof in \cite{Chiachi}, developed for
the $2$-user BC with i.i.d. Rayleigh fading, requires that if
any two subsets (of equal cardinality) are chosen from the set of all rows of
all users' channel matrices, then the joint distribution of the (independent)
rows of one subset is identical to that of the rows of the other subset.
This property holds (after incorporating Step I of proof of Theorem \ref{thm: dof region BC})
when the channel matrices follow a
distribution of type either $\mathcal{D}_0$ or $\mathcal{D}_1$, but not $\mathcal{D}_2$.
The reason is that, under $\mathcal{D}_2$, different rows of a given channel matrix need
not be independent of each other (this is true not just because their norms are
dependent but their directions themselves can be dependent). Hence, the joint
distribution of the set of say, two rows, where both rows are chosen
from the same channel matrix (assuming it has two rows) would be different from
that of another set of two rows where each row comes from channel matrices of different users.
For example, if we consider isotropic fading, the two rows of a given channel
matrix are not independent whereas if the two rows are chosen such that each
comes from a different channel matrix, they can be independent.
Thus, when the channel matrices follow a distribution of type $\mathcal{D}_2$,
even if the BC is degraded after applying Step I of proof of Theorem \ref{thm: dof region BC}, the applicability of the technique of \cite{Chiachi} is not clear; nevertheless, as proved earlier, the technique developed here (more
specifically, Step II.b of the proof of Theorem \ref{thm: dof region BC} which does not require the strong condition that the above alternative approach requires) gives us the DoF region.
\end{remark}

\section{The Two-User MIMO IC}

In this section, we consider a MIMO network with distributed transmitters.
In particular, we consider the problem of characterizing the no-CSIT DoF region of the
two-user MIMO IC with an arbitrary number of antennas at each of four nodes.

\subsection{Channel Model} \label{subsec: Channel Model IC}

\begin{figure} \centering
\includegraphics[height=2.5in,width=3.75in]{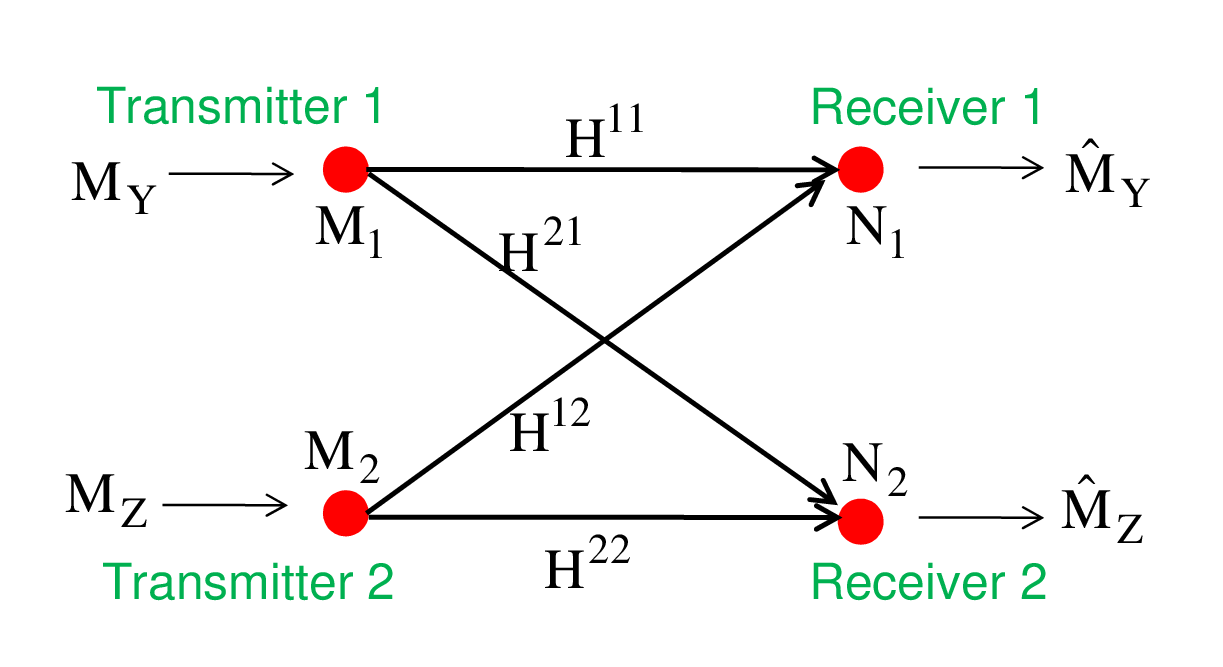}
\caption{The $2$-User MIMO IC.} \label{fig: MIMO_2user_IC}
\end{figure}

Consider the two-user MIMO IC of Fig. \ref{fig: MIMO_2user_IC}
with two transmitter/receiver pairs where ttransmitters $1$ and $2$
have $M_1$ and $M_2$ antennas, respectively, and their corresponding
receivers $1$ and $2$, have $N_1$ and $N_2$ antennas, respectively.
A given transmitter has a message
only for its respective or paired receiver. However, its signal is
received at the unintended receiver as interference. The
input-output relationship is given by
\begin{eqnarray}
\mbox{Receiver 1: } \hspace{1pt} Y(t) = H^{11}(t) X^1(t) + H^{12}(t) X^2(t) + W(t), \\
\mbox{Receiver 2: } Z(t) = H^{21}(t) X^1(t) + H^{22}(t) X^2(t) +
W'(t),
\end{eqnarray}
where at the $t^{th}$ channel use, $Y(t)$ and $Z(t)$ are the
received signals; $X^1(t)$ and $X^2(t)$ are the transmit signals;
$W(t)$ and $W'(t)$ are the additive noises; $H^{11}(t) \in
\mathbb{C}^{N_1 \times M_1}$, $H^{12}(t) \in \mathbb{C}^{N_1 \times
M_2}$, $H^{21}(t) \in \mathbb{C}^{N_2 \times M_1}$, and $H^{22}(t)
\in \mathbb{C}^{N_2 \times M_2}$ are the direct and cross channel
matrices; and there is a power constraint of $P$ at both transmitters.
We assume that all the channel matrices are perfectly and instantaneously
known at both receivers (perfect CSIR). However, the
transmitters know only the distribution of channel matrices (no
CSIT). Let the additive noise be AWGN.

We first consider the model where the channel matrices $H^{11}(t)$ and $H^{21}(t)$ follow a
distribution of type $\mathcal{D}_0(M_1,\bar{N})$ (recall definition \ref{defn:do})
with $\bar{N} = (N_1,N_2)$, whereas the matrices $H^{12}(t)$ and $H^{22}(t)$ follow
a distribution of type $\mathcal{D}_0(M_2,\bar{N})$. Our results are
applicable to a wider class of distributions and these
generalizations are stated in Section \ref{subsec: IC
generalizations}.

Let $M_Y$ and $M_Z$ be the independent messages intended for
receivers $1$ and $2$, respectively. Define the achievability of the
rate pair $(R_1,R_2)$ in the usual way (see the corresponding
definition in the case of BC). The capacity region $\mathcal{C}(P)$
is the set of all achievable rate pairs when the power constraint is
$P$. The DoF region is then defined (as in Definition \ref{defn:dof}) as
\[ \mathbf{D} = \left\{ (d_1,d_2) \left| d_1, d_2 \geq 0 \mbox{ and }
\exists \left(R_1(P),R_2(P)\right) \in C(P) \mbox{ such that } d_i =
\mathrm{MG}(R_i), i = 1,2 \right. \right\}. \]
Further, the transmit signal $X^1(t)$ is independent
of the message $M_Z$ and vice versa. Also both $X^1(t)$ and $X^2(t)$
are independent of the channel matrices and the additive noises.

\subsection{The Inner and Outer Bounds to the DoF Region}

\begin{theorem} \label{thm: inner bound IC}
The inner-bound to the DoF region of the IC with no CSIT is given by
\begin{eqnarray}
\mathbf{D}_{\mathrm{inner}} = \left\{ (d_1,d_2) \left| ~ d_1, ~ d_2
\geq 0, d_1 \leq \min(M_1,N_1), ~ d_2 \leq \min(M_2,N_2), \right.
\frac{d_1}{d_1^*} +\frac{d_2}{d_2^*} \leq 1 \right\}, \label{eq: IC
inner bound}
\end{eqnarray}
where $d_1^*$ and $d_2^*$ are positive numbers such that the
line $\frac{d_1}{d_1^*} + \frac{d_2}{d_2^*} =1$ passes through
points $P_1$ and $P_2$ defined as
\begin{eqnarray}
P_1 \define \left(\min(M_1,N_1), \min\left\{N_2,N_1 -
\left((N_1-M_1)^+ - M_2\right)^+\right\} - \min(N_2,N_1,M_1)\right), \nonumber \\
P_2 \define \left( \min\left\{N_1,N_2 - \left( (N_2-M_2)^+ -
M_1\right)^+\right\} - \min(N_1,N_2,M_2) , \min(N_2,M_2) \right).
\label{def: IC inner bound}
\end{eqnarray}
where for $a, ~ b \in \mathbb{R}$, $(a-b)^+ \define \max(0,a-b)$.
\end{theorem}
\begin{IEEEproof}
See Section \ref{subsec: IC inner-bound proof}.
\end{IEEEproof}

In Fig. \ref{fig: IC typical shape}, we plot the typical shape of
$\mathbf{D}_{\mathrm{inner}}$. Note that depending on the relative
values of $M_1$, $M_2$, $N_1$ and $N_2$, it is possible that
$P_1$ is on the $d_1$-axis and/or $P_2$ is on the $d_2$-axis.
We refer to the bound $\frac{d_1}{d_1^*} +\frac{d_2}{d_2^*} \leq 1$
henceforth as the `inner-bound on the weighted sum'.

\begin{figure} \centering
\includegraphics[height=3.5in,width=4in]{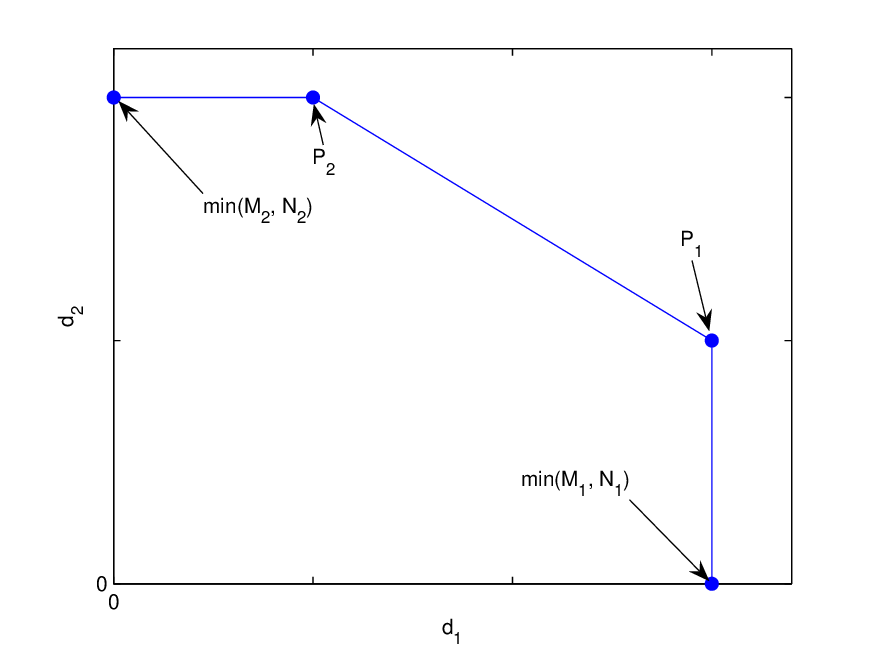}
\caption{The inner-bound for the IC: typical shape} \label{fig: IC
typical shape}
\end{figure}

\begin{remark}[Achievability with respective CSIR]
The above inner-bound is based only on receive zero-forcing and time
sharing. Hence, it is achievable with respective CSIR as well.
\end{remark}

The following theorem gives the outer-bound to DoF region.
\begin{theorem} \label{thm: outer bound IC}
Consider the MIMO IC with no CSIT. Let the channel matrices $H^{11}(t)$
and $H^{21}(t)$ follow a distribution of type
$\mathcal{D}_0(M_1,\bar{N})$ with $\bar{N} = (N_1,N_2)$, whereas the
matrices $H^{12}(t)$ and $H^{22}(t)$ follow a distribution of type
$\mathcal{D}_0(M_2,\bar{N})$. Assume the noise to be AWGN. If $N_1
\geq N_2$, the outer-bound to the DoF region is given by
\begin{eqnarray}
\mathbf{D}_{\mathrm{outer}} = \biggl\{ (d_1,d_2) \left| d_1,~ d_2
\geq 0, ~ d_1 \leq \min(N_1, M_1), ~ d_2 \leq \min(N_2, M_2), \right. \biggr.  \nonumber \\
\left. \frac{d_1}{\min(N_1, M_1)} + \frac{d_2}{\min(N_2, M_1)} \leq
\frac{\min(N_2, M_1+M_2)}{\min(N_2, M_1)} \right\}. \label{eq: IC
outer bound}
\end{eqnarray}
For the case of $N_2 > N_1$, the DoF region can be obtained by
reversing the ordering of the users.
\end{theorem}
\begin{IEEEproof}
See Section \ref{subsec: IC outer-bound proof}.
\end{IEEEproof}
We refer to the above bound on the weighted sum of $d_1$ and $d_2$
as the `outer-bound on the weighted sum'.

\begin{remark}[Comparison of Inner and Outer Bounds]
The inner and outer bounds can be seen to coincide for all values of
the 4-tuple $(M_1,N_1,M_2,N_2)$ except when the condition $\min(N_1,
M_1) > N_2 > M_2$ or its symmetric counterpart $\min(N_2,M_2) > N_1 > M_1$ holds\footnote{It was recently shown
in \cite{Zhu_Guo_noCSIT_DoF_2010} that the inner bound of Theorem \ref{thm: inner bound IC} is in fact tight even in these cases. See Section \ref{subsec: IC open cases} for a further discussion.}.
\end{remark}

\subsection{DoF-Separability}

With perfect CSIT and with fixed channel matrices the DoF region of the MIMO IC
was obtained in \cite{Jafar-Maralle}. It was shown that the perfect CSIT DoF region is
\begin{eqnarray}
\mathbf{D}^{csit} = \Big\{ (d_1,d_2) \left| d_1, d_2 \geq 0, d_1
\leq \min(M_1, N_1), d_2 \leq \min(M_2, N_2), \right. \Big. \nonumber \\
d_1 + d_2 \leq \min \Big. \big\{ M_1 + M_2, N_1+N_2, \max(M_1,N_2),
\max(M_2,N_1) \big\} \Big\}. \label{eq: dof region IC perfect}
\end{eqnarray}
It is natural to ask whether the DoF region is strictly bigger
in the time-varying fading channel setting wherein the fading matrices are random
and assumed to be i.i.d. across channel uses. This
question is interesting in light of the result of \cite{Lalitha_Poor_separability}
where it is shown that the two-user interference channel is not separable, i.e.,
that the capacity of parallel interference channels is higher than that obtained
by separate encoding over the sub-channels and with power allocated optimally
across the sub-channels. Is it possible then that even the perfect-CSIT DoF
region of the IC with i.i.d. fast fading is strictly bigger than the region
defined in equation (\ref{eq: dof region IC perfect})? In Appendix \ref{app: IC separable dof sense},
we answer this question in the negative. In other words, the DoF region of (\ref{eq: dof region IC perfect}) is the DoF region in the i.i.d. fast fading case too, so that the two-user MIMO IC can be said to be {\em DoF-separable}. We are now in a position to compare DoF regions with and without CSIT.



\begin{remark}[Comparison of perfect and no CSIT DoF regions]
\label{rem:perno}
\label{rem: compare dofs of IC} The perfect CSIT and no CSIT DoF regions are
the same if and only if $N_1 \geq N_2 \geq M_1$ or $N_2 \geq N_1 \geq M_2$.
\end{remark}

\begin{remark}
The DoF-optimal transmission scheme presented in
\cite{Jafar-Maralle} (or \cite{Chiachi-Jafar}) for the perfect-CSIT
IC makes use of the null space of the cross channel matrices $H^{12}(t)$ and $H^{21}(t)$. The fact that when $N_1 \geq N_2 \geq M_1$ or $N_2 \geq
N_1 \geq M_2$, perfect-CSIT DoF region can be achieved even without CSIT is not
evident from the achievability scheme of \cite{Jafar-Maralle, Chiachi-Jafar}.
Hence, from these papers, it is not
clear if the no-CSIT DoF region can ever be equal to the perfect-CSIT DoF
region. The study of no-CSIT problem can be seen to yield
CSI-independent robust transmission schemes which achieve perfect-CSIT
DoF regions in cases where this is possible.
\end{remark}

\begin{remark}[The loss of DoF]
When the conditions in Remark \ref{rem:perno} don't hold, the DoF region with perfect
CSIT strictly contains that without CSIT and hence there is a loss of DoF due to lack of CSIT.
For example, the IC with $M_1 = M_2 = 2N$ and $N_1 = N_2 = N$ has sum-DoF of $2N$ with
perfect CSIT and $N$ without CSIT.
\end{remark}

\subsection{Proof of the Inner-Bound} \label{subsec: IC inner-bound proof}
From the shape of the inner-bound shown in Fig. \ref{fig: IC typical
shape}, we observe that it is sufficient to establish the
achievability of points $P_1$ and $P_2$. The achievability of the
whole region then follows by time sharing.

Let us start with point $P_1$. Suppose we want to achieve $d_1 =
\min(M_1,N_1)$, i.e., the maximum DoF
that can be achieved for user $1$. Under this constraint, what is
the maximum DoF that we
can achieve for the second user? Suppose that the second
transmitter sends $d_2$ streams. Due to the complete lack of CSIT,
the transmitters can not employ zero-forcing beam-forming, using
which the achievability of the DoF region under perfect CSIT was
proved in \cite{Jafar-Maralle, Chiachi-Jafar}. As a result, all
$d_2$ streams sent by the second transmitter cause interference at
the first receiver. This receiver zero-forces the interference to
recover the useful signal. Hence, for $d_1 = \min(M_1,N_1)$ to be
achievable for the first user, the second transmitter is constrained
to send at most $\min\{M_2, N_1-\min(M_1,N_1)\} = \min\{M_2,
(N_1-M_1)^+\}$ streams. Then the second receiver receives a total of
$\min(M_1,N_1) + \min(M_2, (N_1-M_1)^+)$ streams, out of which
$\min(M_1,N_1)$ are interference streams for it. Therefore, we can
achieve $d_2 = \min\{ N_2,\min(M_1,N_1) + \min ( M_2, (N_1-M_1)^+)
\} - \min\{N_2,\min(M_1,N_1)\}$, which can be written as $d_2 =
\min\left\{N_2,N_1 - \left( (N_1-M_1)^+ - M_2\right)^+\right\} -
\min(N_2,N_1,M_1)$. This proves the achievability of point $P_1$.
The achievability of point $P_2$ follows by symmetry.

\subsection{Proof of the Outer-Bound} \label{subsec: IC outer-bound proof}
The outer-bound, by definition, is the set of conditions that any
point $(d_1,d_2) \in \mathbf{D}$ must satisfy. Therefore, the
rectangular region defined by $d_1 \leq \min(M_1,N_1)$ and $d_2 \leq
\min(M_2,N_2)$ is a valid outer-bound. The goal of the remaining
part of the proof is to derive the outer-bound on the weighted sum.
Without loss of generality, we may assume that $N_1 \geq N_2$. The
main idea of the proof is similar to the one provided for the BC.

We first enhance the capacity region of the IC by assuming that the
first receiver knows the message $M_Z$. Since the transmit signal
$X^2(t)$ is determined completely by $M_Z$, we may assume that the
first receiver knows $X^2(t)$ as well.

Define $H^1(t) = \begin{bmatrix} H^{11}(t) & H^{12}(t)
\end{bmatrix}$ and analogously $H^2(t)$. Denote by $\mathbf{H}$ the
collection of random variables $\mathbf{H^{11}}$, $\mathbf{H^{12}}$,
$\mathbf{H^{21}}$, and $\mathbf{H^{22}}$.

We now apply Fano's inequality and then take the limit over the
blocklength $n$ to arrive at the following:
\begin{eqnarray}
\lefteqn{ \hspace{1.6cm}  R_2 \leq \frac{1}{n} I
(M_Z;\mathbf{Z}|\mathbf{H}) + \epsilon_n \Longrightarrow R_2 \leq \lim_n
\frac{1}{n} I (M_Z; \mathbf{Z}|\mathbf{H}), }\\
&& {} R_1 \leq \frac{1}{n} I(M_Y; \mathbf{Y}|\mathbf{H}, M_Z) +
\epsilon_n \Longrightarrow R_1 \leq \lim_n \frac{1}{n}
I(M_Y;\mathbf{Y}|\mathbf{H}, M_Z) .
\end{eqnarray}
These bounds are now used to obtain the outer-bound on the weighted
sum. We will again work through the three steps introduced while
proving the DoF region of the BC. The proof is mostly similar and we
will emphasize only the differences.

\emph{\underline{Step I:} } This step is about channel enhancement.
Recall that $H^{11}(t)$ and $H^{21}(t)$ follow a distribution of
type $\mathcal{D}_0(M_1,\bar{N})$, and thus, write $H^{11}(t) =
\Lambda^{11}(t) F^{11}(t)$ and $H^{21}(t) = \Lambda^{21}(t)
F^{21}(t)$. Let $h^1_{\max}(t)$ be the maximum of all diagonal
entries of $\Lambda^{11}(t) $ and $\Lambda^{21}(t) $, and define
$h^{11}_{\max}(t) = h^1_{\max}(t) I_{N_1 \times N_1}$ and
$h^{21}_{\max}(t) = h^1_{\max}(t) I_{N_2 \times N_2}$, where $I_{m
\times m}$, $m>0$, is an $m \times m$ identity matrix.
Note here that all the diagonal entries of matrices
$h^{11}_{\max}(t)$ and $h^{21}_{\max}(t)$ equal $h^1_{\max}(t)$;
these matrices differ only in their sizes. Then, at this
step, define
\begin{eqnarray}
\tilde{Y}(t) &=& h^{11}_{\max}(t) \big( \Lambda^{11}(t) \big)^{-1}
H^1(t) \begin{bmatrix} X^1(t) \\ X^2(t) \end{bmatrix} + W(t)
\label{eq:IC channel enhanced outputs1}\\
\tilde{Z}(t) &=& h_{\max}^{21}(t) \big( \Lambda^{21}(t) \big)^{-1}
H^2(t) \begin{bmatrix} X^1(t) \\ X^2(t) \end{bmatrix} + W'(t).
\label{eq:IC channel enhanced outputs2}
\end{eqnarray}
Then, the analysis of Step I performed in the context of the MIMO BC
implies that
\begin{equation}
I(M_Z;\mathbf{Z}|\mathbf{H}) \leq I(M_Z;\mathbf{\tilde{Z}} |
\mathbf{H}) \mbox{ and } I(M_Y;\mathbf{Y} | M_Z, \mathbf{H}) \leq
I(M_Y;\mathbf{\tilde{Y}} | M_Z, \mathbf{H}). \label{eq: channel
enhancement IC}
\end{equation}
Then the bounds on $d_1$ and $d_2$ are given by
\begin{eqnarray}
\lefteqn{ \hspace{0.7cm} d_2 \leq \underbrace{ \mathrm{MG} \left\{
\lim_n \frac{1}{n} h(\mathbf{\tilde{Z}}| \mathbf{H}) \right\}
}_{(1)} - \underbrace{ \mathrm{MG} \left\{ \lim_n \frac{1}{n}
h(\mathbf{\tilde{Z}}|\mathbf{H},M_Z) \right\} }_{(2)}, \mbox{ and} } \nonumber \\
&& {} d_1 \leq \underbrace{ \mathrm{MG} \left\{ \lim_n \frac{1}{n}
h(\mathbf{\tilde{Y}}|\mathbf{H},M_Z) \right\} }_{(3)} - \underbrace{
\mathrm{MG} \left\{ \lim_n \frac{1}{n} h(\mathbf{\tilde{Y}}|
\mathbf{H}, M_Z, M_Y) \right\} }_{(4)}. \label{eq: IC bounds after
Step I}
\end{eqnarray}

\emph{\underline{Step II:} } A lemma that relates the multiplexing
gains of terms (2) and (3) of the above equations is proved below.
In Step III, we obtain bounds on terms (1) and (4), and then finish the proof of Theorem \ref{thm: outer bound IC}.
\begin{lemma} \label{lem: IC main inequality}
The following inequality holds:
\begin{equation}
\label{eq: IC main inequality}
 \frac{1}{\min(M_1,N_2)} \mathrm{MG} \left\{ \lim_n
\frac{1}{n} h(\mathbf{\tilde{Z}}|\mathbf{H},M_Z) \right\} \geq
\frac{1}{\min(M_1,N_1)} \mathrm{MG} \left\{ \lim_n \frac{1}{n}
h(\mathbf{\tilde{Y}}|\mathbf{H},M_Z) \right\}.
\end{equation}
\end{lemma}
\begin{IEEEproof}
Note that both differential entropy terms in the above inequality
are conditioned on $M_Z$. Given $M_Z$, $\mathbf{X^2}$ is
deterministic. Hence
\[
h \Big( \mathbf{\tilde{Z}} \Big| M_Z, \mathbf{H} \Big) = h
\Big(\mathbf{\tilde{Z}} - \mathbf{h_{\max}^{21}} \big(
\mathbf{\Lambda^{21}} \big)^{-1} \mathbf{H^{22}} \mathbf{X^2} \Big|
\mathbf{H} \Big)
\]
because translation does not change differential entropy and
\[
\mathbf{\tilde{Z}} - \mathbf{h_{\max}^{21}} \big(
\mathbf{\Lambda^{21}} \big)^{-1}\mathbf{H^{22}} \mathbf{X^2} =
\mathbf{h_{\max}^{21}} \mathbf{F^{21}} \mathbf{X^1} + \mathbf{W'}
\define \mathbf{\tilde{Z}}'
\]
is independent of $M_Z$. Also define
\[
\mathbf{\tilde{Y}} - \mathbf{h_{\max}^{11}} \big(
\mathbf{\Lambda^{11}} \big)^{-1} \mathbf{H^{12}} \mathbf{X^2} =
\mathbf{h_{\max}^{11}} \mathbf{F^{11}} \mathbf{X^1} + \mathbf{W}
\define \mathbf{\tilde{Y}}'.
\]
Hence it is sufficient to prove the inequality of the lemma with
$\mathbf{\tilde{Y}}$ and $\mathbf{\tilde{Z}}$ replaced by
$\mathbf{\tilde{Y}}'$ and $\mathbf{\tilde{Z}}'$, respectively.

After having eliminated the signal $\mathbf{X^2}$, we are left with
only the first transmitter and the two receivers (i.e., a BC with $M_1$ transmit
antennas and two receivers with $N_1$ and $N_2$ antennas).
Moreover, conditioned on $\mathbf{H}$ and
$M_Z$, the signals $\mathbf{\tilde{Y}}$ and $\mathbf{\tilde{Z}}$ are
statistically equivalent in the following sense. For a given integer
$m$ such that $0 < m \leq \min(N_1,N_2)$, the joint distribution,
conditioned on $\mathbf{H}$ and $M_Z$, of any $m$ random variables chosen from the
set of $N_1$ random variables $\{ \mathbf{\tilde{Y}}_i\}_{i=1}^{N_1}$ is
identical to that of any $m$ random variables chosen from the set of $N_2$ random variables
$\{ \mathbf{\tilde{Z}}_i\}_{i=1}^{N_1}$ (this property is referred
in the sequel as the statistical equivalence of $\mathbf{\tilde{Y}}$
and $\mathbf{\tilde{Z}}$). Hence, the arguments in the proof of
Lemma \ref{lem: main ineq}, developed for the BC, can be directly
applied to obtain the inequality of the present lemma.

\end{IEEEproof}

\emph{\underline{Step III:} } Finally, terms (1) and (4) in equation (\ref{eq: IC bounds after
Step I}) are easily bounded as
\begin{equation}
\mathrm{MG} \left\{ \lim_n
\frac{1}{n} h(\mathbf{\tilde{Z}}| \mathbf{H}) \right\} \leq
\min(M_1+M_2,N_2) \quad {\rm and} \quad \mathrm{MG} \left\{ \lim_n
\frac{1}{n} h(\mathbf{\tilde{Y}}| \mathbf{H}, M_Z,M_Y) \right\} = 0
\end{equation}

Using these two facts, Lemma \ref{lem: IC main inequality}, and
the bounds given in equation (\ref{eq: IC bounds after Step I}), we
obtain the outer bound on the weighted sum, concluding the proof of
Theorem \ref{thm: outer bound IC}.

\subsection{Discussion of the Cases where the Inner and Outer Bounds
Do Not Coincide} \label{subsec: IC open cases}

\begin{figure} \centering
\includegraphics[height=4in,width=5in]{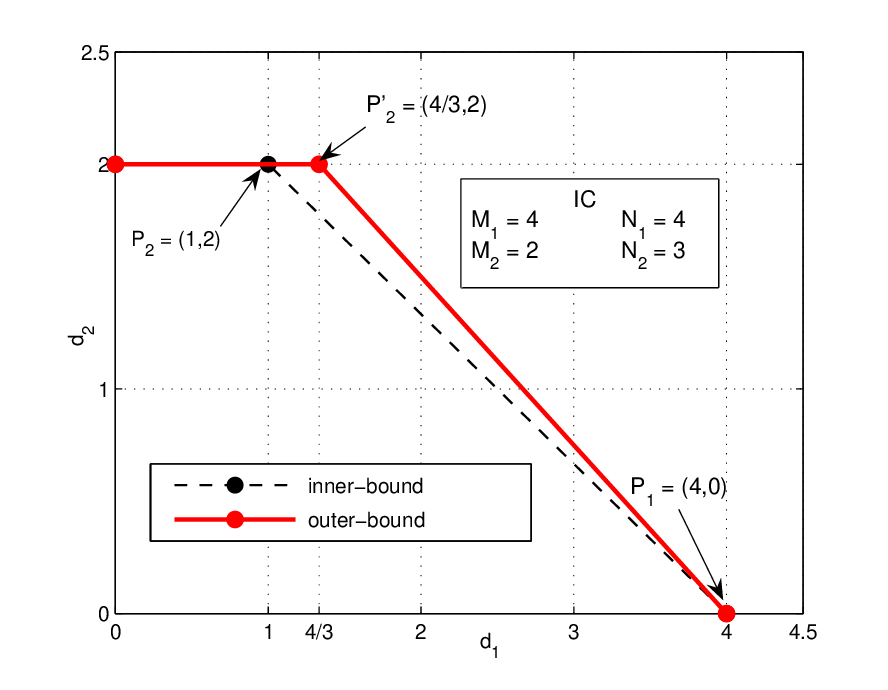}
\caption{The inner and outer bounds for the IC with
$(M_1,N_1,M_2,N_2) = (4,4,2,3)$.} \label{fig: IC example bounds do
not meet}
\end{figure}

Consider, without loss of generality, the case of $\min(M_1,N_1) >
N_2 > M_2$ where the inner and outer bounds don't coincide. An example of the IC that falls under this category is
considered in Fig. \ref{fig: IC example bounds do not meet},
where the inner and outer-bounds to the DoF region of the IC with
$(M_1,N_1,M_2,N_2) = (4,4,2,3)$ are plotted. The inner-bound on the
weighted sum passes through a point $P_2 = (1,2)$, whereas the
outer-bound passes through a point $P_2' = (\frac{4}{3},2)$. To
achieve any point lying on the line segment joining $P_2$ and $P_2'$
(not including $P_2$), the first transmitter must send more than one
stream. However, these streams should span only a $1$-dimensional
subspace at the second receiver, because, otherwise, we can not
achieve $2$ DoF for the second user. This is possible only if the
first transmitter can exploit the null space of $H^{21}(t)$, the
channel matrix between the first transmitter and the second
receiver. However, due to the absence of CSIT, this is not feasible
and point $P_2'$ is not achievable. Indeed, it has been proved
recently that the outer-bound derived in Theorem \ref{thm: outer
bound IC} is loose and the DoF region of the no-CSIT IC is equal to
the inner-bound stated in Theorem \ref{thm: inner bound IC}
\cite{Zhu_Guo_noCSIT_DoF_2010}.

However, using the idea of blind interference alignment
in \cite{Jafar_correlations}, the point
$P_2'$ can be shown to be achievable without CSIT\footnote{In \cite{Jafar_correlations}
the example of the IC with $(M_1,N_1,M_2,N_2) = (1,2,3,4)$ was considered.} (i.e., even
when the transmitters do not know the null spaces of the channel
matrices) for a staggered block fading model (to which Theorem
\ref{thm: outer bound IC} does not apply). In that achievability scheme, the transmitter(s)
makes use of the fact that the null spaces of the required channel
matrices remain constant over the coherence period thereby allowing the
specification of the beam-forming vectors/matrices that minimize the dimension of
the interference-subspace at the receiver(s) to
achieve point $P_2'$. 

\subsection{Generalization} \label{subsec: IC generalizations}

The following generalizations of the above result are possible.

\begin{theorem} \label{thm: IC generalization}
Consider the IC with no CSIT. If for a given $j \in
\{1,2\}$, the channel matrices $H^{1i}(t)$ and $H^{2i}(t)$ follow a
distribution of type $\mathcal{D}_j (M_i,N_1,N_2)$ $\forall$ $i \in
\{1,2\}$, then the inner and outer bounds to the DoF region of the
no-CSIT IC with AWGN are given by those defined by
equations (\ref{eq: IC inner bound}) of Theorem \ref{thm: inner
bound IC} and (\ref{eq: IC outer bound}) of Theorem \ref{thm: outer
bound IC}, respectively.
\end{theorem}
\begin{IEEEproof}
See Appendix \ref{app: proof of thm: IC generalization}.
\end{IEEEproof}

\begin{remark} The DoF regions of the 2-user MIMO
for the subclasses $ M_1 \leq N_2, M_2 \leq N_1 $
and $ M_1 \geq N_1, M_2 \geq N_2 $ were previously obtained
for the i.i.d. Rayleigh fading model in \cite{Chiachi}.
Moreover, as mentioned in the introduction, under more restrictive assumptions on
the distributions of the channel matrices and additive noises, (the
same) inner and outer bounds to the DoF region of the no-CSIT IC
were also derived independently in \cite{Chiachi2, D.Guo}. In
particular, the case of i.i.d. Rayleigh fading was considered in
\cite{Chiachi2} and the more general case of isotropic fading was
considered in \cite{D.Guo}. To obtain the
outer-bound, \cite{D.Guo} needs a crucial lemma, a
counterpart of Lemma \ref{lem: IC main inequality},
which is proved therein using the ``generalized
super-additive property of differential entropy"
which in turn critically depends on the
assumption of isotropic fading. The proof here does not
make use of such a property and also yields a more general
result.
\end{remark}

\section{The Cognitive Radio Channel} \label{sec: CRC}

\subsection{Channel Model} \label{subsec: CRC channel model}

\begin{figure} \centering
\includegraphics[height=2.5in,width=3.75in]{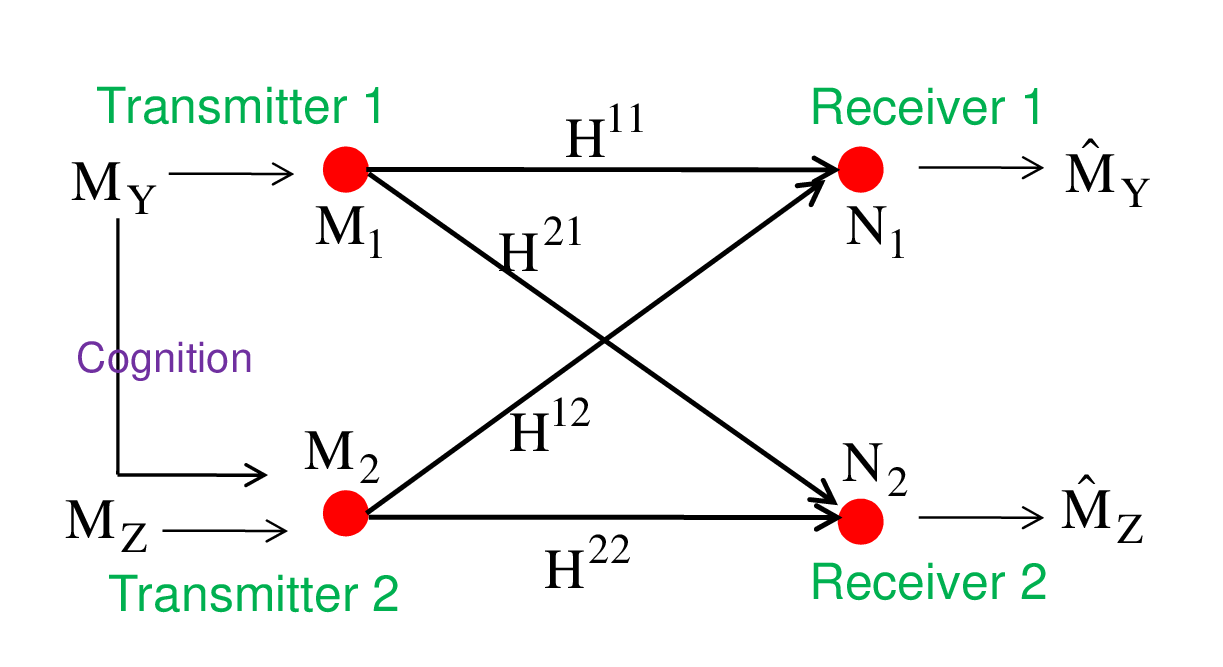}
\caption{The $2$-User MIMO CRC.} \label{fig: MIMO_2user_CRC}
\end{figure}

The input-output relationship of the CRC is same as that of the MIMO IC.
It differs from the IC because of one assumption, which is that one of the transmitters (here, the
second transmitter, also called the cognitive transmitter) knows the message of
the other `primary' transmitter non-causally. See Fig. \ref{fig: MIMO_2user_CRC}. The receiver of the cognitive
transmitter (CT) is called the cognitive receiver (CR) while the
other transmit-receive pair is called the primary pair (denoted as PT and
PR, respectively).

Because of the non-causal knowledge of the message of the PT at the CT,
the CT, besides transmitting its own message, can also aid the PT to transmit its message.

Since the channel model of CRC is the same as that of the IC except
for one extra assumption, all the definitions in \ref{subsec:
Channel Model IC} apply. However, we assume that the channel matrices
$H^1(t)$ and $H^2(t)$ follow a distribution of type
$\mathcal{D}_0(M_1 + M_2,N_1,N_2)$ and that the noise is AWGN.

\subsection{The Inner and Outer Bounds}

The inner-bound is given by the following theorem.
\begin{theorem} \label{thm: inner bound CRC}
The inner-bound to the DoF region of the CRC with AWGN under the no CSIT assumption when the channel matrices
$H^1(t)$ and $H^2(t)$ follow a distribution of type
$\mathcal{D}_0(M_1 + M_2,N_1,N_2)$
is given as
\[ \mathbf{D}_{\mathrm{inner}} = \left\{ (d_1, d_2) \left| d_1,d_2
\geq 0, d_2 \leq \min(M_2,N_2), \frac{d_1}{d_1^*} +
\frac{d_2}{d_2^*} \leq 1 \right. \right\}, \]
where $d_1^*$ and $d_2^*$ are such that the line
$\frac{d_1}{d_1^*} + \frac{d_2}{d_2^*} = 1$ passes through points
$P_1$ and $P_2$ which in turn are defined as
\begin{eqnarray*}
\lefteqn{ \hspace{2mm} P_1 = (\min(N_1,M_1+M_2),0) }\\
&& {} \hspace{-5mm} P_2 = \left( \min\left\{N_1,N_2 - \left(
(N_2-M_2)^+ - M_1\right)^+\right\} - \min(N_1,N_2,M_2) ,
\min(N_2,M_2) \right).
\end{eqnarray*}
\end{theorem}
\begin{IEEEproof}
See Section \ref{Sec:crcib}.
\end{IEEEproof}

\begin{figure} \centering
\includegraphics[height=3.1in,width=3.5in]{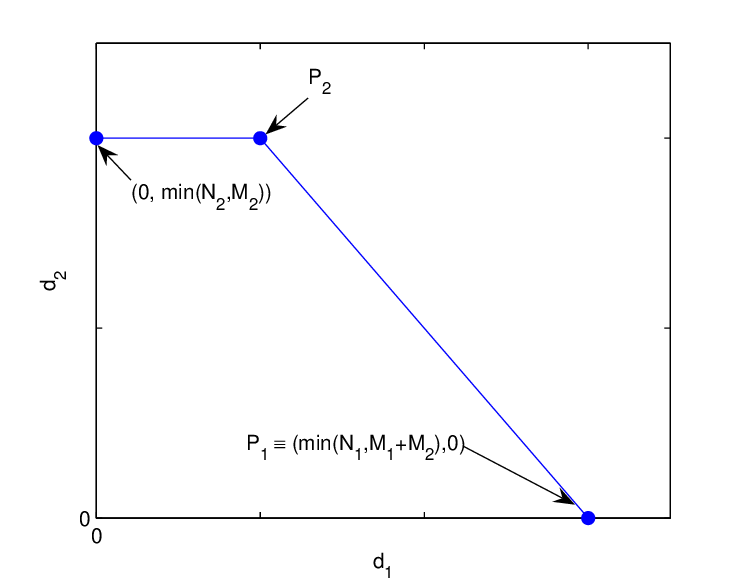}
\caption{CRC: typical shape of the inner-bound} \label{fig: typical
shape CRC}
\end{figure}

The typical shape of
$\mathbf{D}_{\mathrm{inner}}$ is shown in Fig. \ref{fig: typical shape CRC}. Note that point $P_2$ can be on
$d_2$-axis.

The outer bound is stated next.
\begin{theorem} \label{thm: outer-bound CRC}
The outer-bound to the DoF region of the no-CSIT CRC, when $H^1(t)$
and $H^2(t)$ follow a distribution of type
$\mathcal{D}_0(M_1+M_2,N_1,N_2)$ and noise is AWGN, is given by
\begin{eqnarray*}
\lefteqn{ \mathbf{D}_{\mathrm{outer}} = \biggl\{ (d_1,d_2) \left|
d_1, d_2 \geq 0, ~ d_1 \leq \min(N_1, M_1+M_2), ~ d_2 \leq
\min(N_2, M_2), \right. \biggr. } \\
&& {} \hspace{4.5mm} \left. \frac{d_1}{\min(N_1, M_1 + M_2)} +
\frac{d_2}{\min(N_2, M_1+ M_2)} \leq 1 \hspace{10pt}  \hspace{7.5mm}
\hspace{1pt} \cdots ~ \mbox{if } N_1 \geq N_2 \right. \\
&& {} \left. \frac{d_1}{\min(N_1, M_2)} + \frac{d_2}{\min(N_2, M_2)}
\leq \frac{\min(N_1, M_1+M_2)}{\min(N_1,M_2)} \hspace{10pt} ~ \cdots
~ \mbox{if } N_1 < N_2 \right\}.
\end{eqnarray*}
\end{theorem}
\begin{IEEEproof}
See Section \ref{Sec:crcob}.
\end{IEEEproof}

\begin{remark}[Comparison of inner and outer bounds]
We observe that the inner and outer bounds coincide and give us the
exact characterization of the DoF region, except if $\min(N_1,
M_1+M_2)
> N_2 > M_2$. 
\end{remark}

\begin{remark}[Perfect-CSIT DoF region]
The DoF region of the CRC with perfect CSIT is given
by\footnote{Again, \cite{Chiachi-Jafar} deals with the case of
deterministic (fixed) channel matrices, whereas we are dealing here
with the case of time-varying channel matrices. However, the
converse of \cite{Chiachi-Jafar} and the discussion in Appendix
\ref{app: IC separable dof sense} imply that the perfect-CSIT DoF
region of the CRC considered here equals the region stated in this
remark. In other words, the MIMO CRC is also DoF-separable.} \cite{Chiachi-Jafar}
\begin{eqnarray}
\mathbf{D}^{csit} = \Big\{ (d_1,d_2) \left| d_1, d_2 \geq 0, d_1
\leq \min(M_1+M_2,N_1), d_2 \leq \min(M_2,N_2), \right. \Big. \nonumber \\
d_1 + d_2 \leq \min \Big. \left\{ M_1 + M_2, N_1+N_2, \max(M_2,N_1)
\right\} \Big\}.
\end{eqnarray}
\end{remark}

\begin{remark}[Comparison of perfect and no CSIT DoF regions of the CRC]
The DoF regions of the CRC with perfect and no CSIT are identical
only if $N_1 > N_2 \geq M_1 + M_2$ or $N_2 \geq N_1 \geq M_2$. Note that if $N_2 \geq N_1 \geq M_2$, it is still possible that $M_1 + M_2 > N_2$ in which case the perfect-CSIT DoF-region optimal transmission scheme presented in \cite{Chiachi-Jafar} will still make use of the null space of channel matrix $\begin{bmatrix} H_{21}(t) & H_{22}(t) \end{bmatrix}$. However, from the results of Theorems \ref{thm: inner bound CRC} and \ref{thm: outer-bound CRC}, we know that whenever $N_2 \geq N_1 \geq M_2$, CSIT is not necessary for DoF-optimal performance. Hence, for the case of $N_2 \geq N_1 \geq M_2$, the achievability scheme in this paper improves upon the CSI-dependent scheme of \cite{Chiachi-Jafar} in that it achieves the same DoF performance without CSIT.
\end{remark}

\begin{remark}[Comparison of the DoF regions of the IC and the CRC]
Let us now determine when it is useful, in terms of the DoF region,
to have a cognitive transmitter. If $N_1 > M_1$, the DoF region of
the CRC is always bigger than that of the IC, because the maximum
number of DoF achievable for the first user increase from $M_1$ to
$\min(N_1,M_1+M_2) > M_1$. Consider now the case of $N_1 \leq M_1$
wherein the maximum number of DoF achievable for the first user do
not increase when the second transmitter is made cognitive. It is
easy to verify that, under no CSIT, the inner bound remains
unchanged in going from the IC to the CRC. But, quite interestingly,
when there is perfect CSIT, the DoF region of the CRC is, in
general, strictly bigger than that of the corresponding IC. To
understand this, consider the following. When there is perfect CSIT,
the dimension of the interference subspace at the second receiver is
equal to the number of streams intended for the first receiver minus
the dimension of the null spaces of $H_{21}(t)$ and $\begin{bmatrix}
H_{21}(t) & H_{22}(t) \end{bmatrix}$ in the cases of
the IC and the CRC, respectively,  because the signal intended for the first
receiver can be transmitted by only the first transmitter in the
case of the IC whereas it can be done by both transmitters in the
case of the CRC. Since the dimension of the null space of
$\begin{bmatrix} H_{21}(t) & H_{22}(t) \end{bmatrix}$ is in general
higher than that of $H_{21}(t)$, making the second transmitter
cognitive helps in reducing the interference at the second receiver.
However, this reduction in the interference at the second
transmitter is feasible only with perfect CSIT, and hence, under no
CSIT, the two inner bounds are equal.

In summary, it is useful to
have the second transmitter cognitive only if $N_1 > M_1$
given currently known inner-bounds.
\end{remark}


A possible generalization of the model of the CRC is to allow the
possibility of having one or more terminals cognitive at the same
time. The DoF regions of such channels were derived in
\cite{Chiachi-Jafar} for the case of perfect CSIT. It turns out that
the general techniques developed here are useful for characterizing
its no-CSIT DoF region as well. See
\cite{Vaze_Dof_Cognitive_IC_ISIT} for details.

\subsection{Proof of the Inner-Bound}
\label{Sec:crcib}
It is sufficient to prove the achievability of points $P_1$ and
$P_2$. Let us start with point $P_1$. Since the CT knows the message
to be transmitted by the PT, the maximum DoF that are achievable for
the primary pair are $d_1 = \min(N_1, M_1+M_2)$. Now, when $d_1 =
\min(N_1, M_1+M_2)$, we do not know how to achieve any positive DoF
for the CT-CR pair, i.e., $d_2 = 0$. This is because depending upon
the relative values of $N_1$ and $M_1+M_2$, either the CT uses all
$M_2$ streams or all possible $N_1$ DoF of the received signal-space
of the PR are used up.

Consider now point $P_2$. Note that this point is the same as the
corresponding point defined for the IC (cf. equation (\ref{def: IC
inner bound})). That is, given $d_2 = \min(M_2,N_2)$, the maximum
DoF known to be achievable for the first user are
$\min\left\{N_1,N_2 - \left( (N_2-M_2)^+ - M_1\right)^+ \right\} -
\min(N_1,N_2,M_2)$, irrespective of whether the second transmitter
is cognitive or not. A simple argument explains this. If $N_2 \leq
M_2$, all possible DoF available at the second receiver are used up
to achieve $d_2 = \min(M_2,N_2) = N_2$, and therefore, we can not
achieve any positive DoF for the first user. Now, if $M_2 < N_2$,
the second transmitter has used up all available $M_2$ DoF, and
hence, we do not know how to improve the DoF
achievable for the first user, even though the
second transmitter is cognitive.

\subsection{Proof of the Outer-Bound}
\label{Sec:crcob}
Again, the rectangular region defined by the constraints $d_1 \leq
\min(N_1,M_1+M_2)$, and $d_2 \leq \min(M_2,N_2)$ is certainly an
outer-bound. Thus, only the outer-bound on the weighted sum needs to
be derived. We have to consider the two cases, namely, $N_1 \geq
N_2$ and $N_2 > N_1$ separately. Let us begin with the first case.

We apply Fano's inequality assuming that the PR knows the message
$M_Z$. Again, the proof is similar to those provided in the cases of
BC and IC. It consists of three steps.

\emph{\underline{Step I:} } We directly state the bounds on $d_1$
and $d_2$ that we get after the step of channel enhancement. In
fact, when $N_1 \geq N_2$ these bounds are identical to the
corresponding bounds derived for the IC (cf. equation (\ref{eq: IC
bounds after Step I})) .
\begin{eqnarray}
\lefteqn{ \hspace{0.7cm} d_2 \leq \underbrace{ \mathrm{MG} \left\{
\lim_n \frac{1}{n} h(\mathbf{\tilde{Z}}| \mathbf{H}) \right\}
}_{\leq ~ \min(N_2, M_1+M_2)} - \mathrm{MG} \left\{ \lim_n
\frac{1}{n} h(\mathbf{\tilde{Z}}|\mathbf{H},M_Z) \right\}, }
\label{eq: N1 geq N2 stepI 1}\\
&& {} d_1 \leq \mathrm{MG} \left\{ \lim_n \frac{1}{n}
h(\mathbf{\tilde{Y}}|\mathbf{H},M_Z) \right\}  - \underbrace{
\mathrm{MG} \left\{ \lim_n \frac{1}{n}
h(\mathbf{\tilde{Y}}|\mathbf{H},M_Z,M_Y) \right\} }_{= ~ 0},
\label{eq: N1 geq N2 stepI 2}
\end{eqnarray}
where $\mathbf{\tilde{Y}}$ and $\mathbf{\tilde{Z}}$ are as in
equation (\ref{eq: IC bounds after Step I}).

\emph{\underline{Step II:} } Consider the following lemma.
\begin{lemma} \label{lem: CRC N1 geq N2}
The following inequality holds:
\[ \frac{1}{\min(M_1+M_2, N_2)} \mathrm{MG} \left\{ \lim_n
\frac{1}{n} h(\mathbf{\tilde{Z}}|\mathbf{H},M_Z) \right\} \geq
\frac{1}{\min(M_1+M_2, N_1)} \mathrm{MG} \left\{ \lim_n \frac{1}{n}
h(\mathbf{\tilde{Y}}|\mathbf{H},M_Z) \right\}. \]
\end{lemma}
\begin{IEEEproof}
The proof of this lemma is similar to that of Lemma \ref{lem: IC
main inequality}. The important difference here is that since the
second transmitter is cognitive, its signal $X^2(t)$ is determined
by both the messages and not just the message $M_Z$, and therefore,
conditioned on $M_Z$, the signal $\mathbf{X^2}$ of the second
transmitter is not deterministic.

Therefore, as far as this lemma is concerned, we have to consider
the BC with $M_1 + M_2$ transmit antennas (obtained by pooling
the antennas at the two transmitters) and two receivers
with $N_1$ and $N_2$ antennas. Then, applying Lemma \ref{lem: main
ineq}, we obtain the required result.
\end{IEEEproof}

\emph{\underline{Step III:} } This follows in the standard way. This
completes the proof for the first case and let us now consider the
second case.

Here, $N_2 > N_1$. We assume that the CR knows the message $M_Y$ of
the primary pair. Again, we use Fano's inequality. The bounds
obtained at the end of Step I are stated below.

\emph{\underline{Step I:} } We have
\begin{eqnarray*}
\lefteqn{ \hspace{0.7cm} d_1 \leq \underbrace{ \mathrm{MG} \left\{
\lim_n \frac{1}{n} h(\mathbf{\tilde{Y}}| \mathbf{H}) \right\}}_{\leq
\min(N_1, M_1+M_2)}  - \mathrm{MG} \left\{ \lim_n \frac{1}{n}
h(\mathbf{\tilde{Y}}|\mathbf{H},M_Y) \right\}, } \\
&& {} d_2 \leq \mathrm{MG} \left\{ \lim_n \frac{1}{n}
h(\mathbf{\tilde{Z}}|\mathbf{H},M_Y) \right\} - \underbrace{
\mathrm{MG} \left\{ \lim_n \frac{1}{n}
h(\mathbf{\tilde{Z}}|\mathbf{H},M_Z, M_Y) \right\} }_{= 0},
\end{eqnarray*}
where $\mathbf{\tilde{Y}}$ and $\mathbf{\tilde{Z}}$ are as before.

\begin{figure} \centering
\includegraphics[height=4in,width=5in]{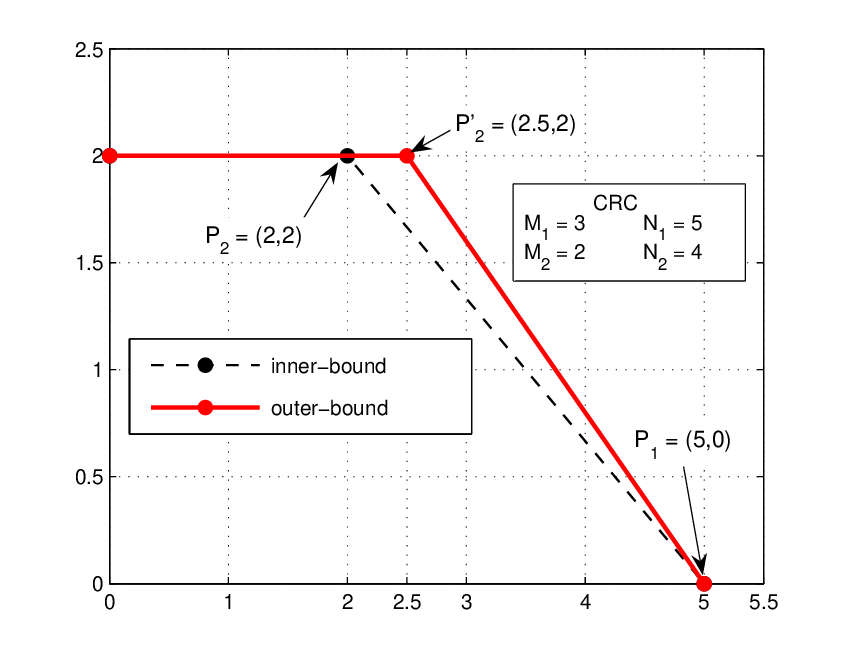}
\caption{The inner and outer bounds for the CRC with
$(M_1,N_1,M_2,N_2) = (3,5,2,4)$.} \label{fig: CRC example bounds
dont meet}
\end{figure}

\emph{\underline{Step II:} }
\begin{lemma}
The following inequality holds:
\[ \frac{1}{\min(M_2, N_1)} \mathrm{MG} \left\{ \lim_n
\frac{1}{n} h(\mathbf{\tilde{Y}}|\mathbf{H},M_Y) \right\} \geq
\frac{1}{\min(M_2, N_2)} \mathrm{MG} \left\{ \lim_n \frac{1}{n}
h(\mathbf{\tilde{Y}}|\mathbf{H},M_Y) \right\}. \]
\end{lemma}
\begin{IEEEproof}
Now, conditioned on $M_Y$, the signal $\mathbf{X^1}$ is
deterministic. Hence, as far as this lemma is concerned, we have the
BC with CT as its transmitter, and PR and CR as its receivers.
\end{IEEEproof}

\emph{\underline{Step III:} } Follows in the standard way. \IEEEQED

\subsection{Discussion of the Case where the Inner and Outer Bounds
Do Not Coincide} \label{subsec: CRC open case}

Consider the example of the CRC given by $(M_1, N_1, M_2, N_2) = (3, 5, 2, 4)$
for which the condition $\min(N_1, M_1+M_2) > N_2 > M_2$ holds. The
inner and outer bounds to the DoF region are shown in Fig. \ref{fig:
CRC example bounds dont meet}. Here, the outer bound on the weighted
sum passes through a point $P_2' = (2.5, 2)$, whereas the inner
bound passes through a point $P_2 = (2,2)$. We believe that to achieve
any point lying on the line segment joining points $P_2$ and $P_2'$
(not including $P_2$), it is necessary that the transmitters are able to
exploit in some manner the null space of the channel matrix to the
second receiver (i.e., of $\begin{bmatrix} H_{21}(t) & H_{22}(t)
\end{bmatrix}$). We therefore conjecture that, as in the case of the IC,
the inner-bound to the no-CSIT DoF region is tight. Thus, a better
bounding technique is needed to derive a tight outer-bound.

\section{Extension to $K$-User MIMO IC, CRC, and the $X$ Networks}
\label{sec:networks}
In this section, we consider important classes of $K$-user interference networks and obtain
their no-CSIT DoF regions in some special cases.

\subsection{The $K$-User IC}

\begin{figure} \centering
\includegraphics[height=3.5in,width=4in]{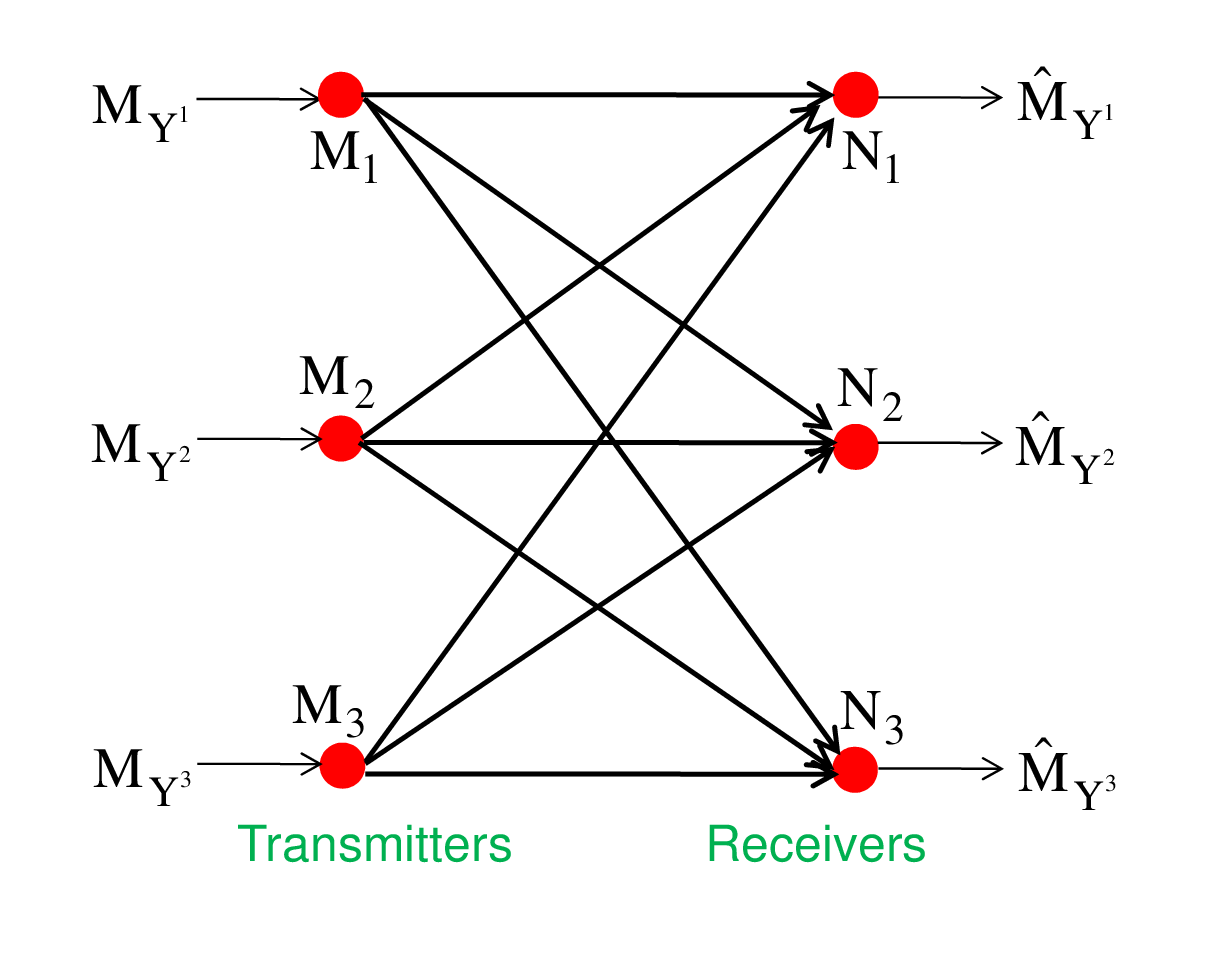}
\caption{The $K$-User MIMO IC with $K = 3$.} \label{fig:
MIMO_Kuser_IC}
\end{figure}

The $K$-user MIMO IC is defined as a generalization of the $2$-user
MIMO IC and is shown in Fig. \ref{fig: MIMO_Kuser_IC} for the case
of $K = 3$. The input-output relationship is given by
\[ Y^i(t) = \sum_{j=1}^K H^{ij}(t) X^j(t) + W^i(t),\]
where $Y^i \in \mathbb{C}^{N_i \times 1}$ is the signal received at
the $i^{th}$ user; $X^j(t) \in \mathbb{C}^{M_j \times 1}$ is the
signal transmitted by transmitter $j$; $H^{ij}(t) \in \mathcal{C}^{N_i \times
M_j}$ is the channel matrix from transmitter $j$ to receiver $i$;
$W^i(t)$ is additive noise. There is a power
constraint of $P$ at all transmitters. It is assumed that all
receivers know all the channel matrices perfectly and
instantaneously, while the transmitters know only their distribution
(i.e., global CSIR and no CSIT).

Let $H^i(t) = \left[ \begin{array}{ccc}
H^{i1}(t) & \cdots & H^{iK}(t) \\
\end{array} \right]$. The channel matrices $\{H^i\}$ follow a
distribution of type $\mathcal{D}_0(\sum_i M_i, \bar{N})$ with
$\bar{N} = (N_1, N_2, \cdots, N_K)$. Further, we take noise to
i.i.d. $\mathcal{C}\mathcal{N}(0,1)$. Lastly, channel and noise
realizations are taken to be i.i.d. across time.

The DoF region is defined in the standard manner.

We have the following results about the DoF region of the $K$-user
IC. Let $by$ $M_{\rm{tot}} = \sum_i M_i$.
\begin{lemma}
The outer-bound to the DoF region of the $K$-user IC with no CSIT is
given by
\begin{equation}
\mathbf{D}^{\rm{outer}} = \left\{ (d_1,\cdots,d_K) | 0 \leq d_i \leq
\min(M_i,N_i), \sum_{i=1}^K \frac{d_i}{\min(N_i, M_{\rm{tot}}) }
\leq 1 \right\}. \label{eq: Kuser IC outer-bound}
\end{equation}
\end{lemma}
\begin{IEEEproof}
For User $i$, the maximum achievable DoF can not exceed
$\min(M_i,N_i)$. If we assume that all the transmitters can
cooperate perfectly, then the DoF region of the resulting BC will be
an outer-bound to the DoF region of the IC. For the resulting BC, by
Theorem \ref{thm: dof region BC}, we observe that the condition
$\sum_{i=1}^K \frac{d_i}{\min(N_i, M_{\rm{tot}}) } \leq 1$ must
hold. We refer to this type of outer-bound as the `overall BC
outer-bound'.
\end{IEEEproof}

This outer-bound stated in the lemma is tight in the following
cases.
\begin{theorem} \label{thm: K-user IC}
The DoF region of the $K$-user MIMO IC with no CSIT is given by the
region defined in equation (\ref{eq: Kuser IC outer-bound}),
provided one of the following conditions hold:
\begin{enumerate}
\item $N_i \leq M_i$ $\forall$ $i$, and
\item $N_i = N$, $M_i = M$ $\forall$ $i$ and $N > M$.
\end{enumerate}
\end{theorem}
\begin{IEEEproof}
We only need to prove the achievability part. Consider Case $1)$.
The outer-bound is defined by $\sum_i \frac{d_i}{N_i}
\leq 1$, and the achievability follows by time division. In Case
$2)$, the outer-bound is defined by conditions $d_i \leq M$
$\forall$ $i$ and $\sum_i d_i \leq \min(N, KM)$. Therefore, the
entire outer-bound is achievable by receive zero-forcing and time
sharing.
\end{IEEEproof}

\begin{remark}
The above theorem provides the complete characterization of the case
where all the transmitters have equal number of antennas, and so do
all receivers. Furthermore, this result is more general than the
corresponding perfect-CSIT result in the sense that the DoF region
of the perfect-CSIT $K$-user MIMO IC is known only if
$\frac{\max(M,N)}{\min(M,N)}$ is an integer.
\end{remark}

\begin{remark}
Recently, it has been proved that over the time-varying $K$-user
SISO IC ($M_i = N_i = 1$, $\forall i$) with perfect channel
knowledge at all nodes, $\frac{K}{2}$ sum-DoF are achievable almost
surely using the technique of interference alignment \cite{Cadambe}.
However, in light of the above theorem, we see that the sum-DoF are
limited to $1$ when there is no CSIT. The work of \cite{Cadambe} has
been generalized in \cite{GouJafar} to the case of time-varying MIMO
IC with perfect CSI at all nodes where all transmitters have $M$
antennas and all receivers have $N$ antennas each. If we compare the
sum-DoF achievable with perfect CSIT and no CSIT in the special case
of $M = N$, we observe that the sum-DoF collapse from $\frac{MK}{2}$
with perfect CSIT to $M$ without CSIT.
\end{remark}

\subsection{The $K$-User CRC}

\begin{figure} \centering
\includegraphics[height=3.5in,width=4in]{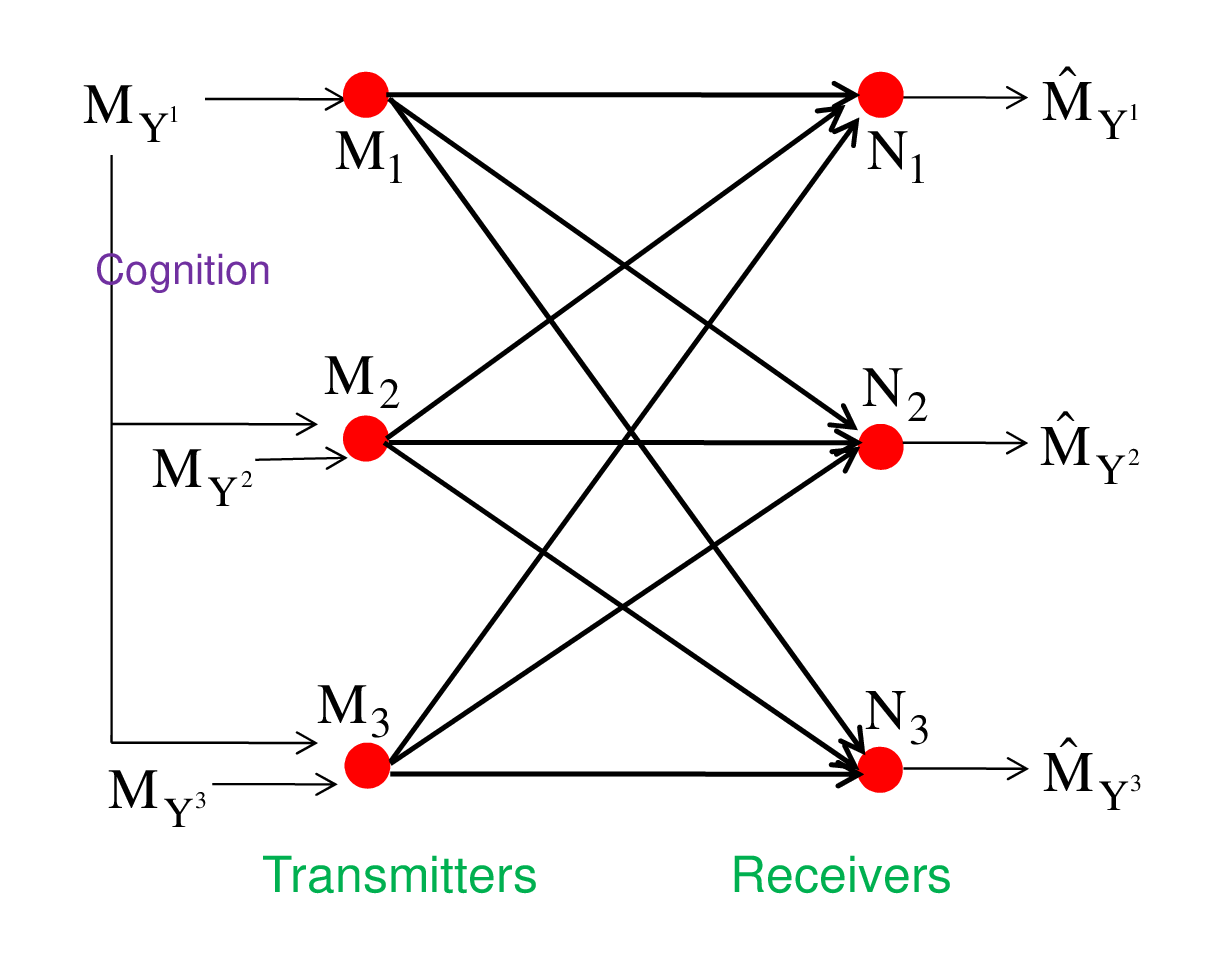}
\caption{The $K$-User MIMO CRC with $K = 3$.} \label{fig:
MIMO_Kuser_CRC}
\end{figure}

We define the $K$-user MIMO CRC as the $K$-user MIMO IC wherein the
first transmit-receive pair is primary while all other pairs are
cognitive, i.e., transmitters $2$ to $K$ know the message of the
primary/first transmitter non-causally. The $K$-user CRC for the
case of $K = 3$ is shown in Fig. \ref{fig: MIMO_Kuser_CRC}. We have
the following results for this CRC.

\begin{lemma}
The outer-bound to the DoF region of the $K$-user CRC with no CSIT
is given by
\begin{eqnarray}
\lefteqn{ \hspace{-4cm}  \mathbf{D}^{\rm{outer}} = \Big\{
(d_1,\cdots,d_K) | 0 \leq d_1 \leq \min(M_{\rm{tot}}, N_1), ~
0 \leq d_i \leq \min(M_i, N_i) ~ \forall ~ i > 1, \Big. } \nonumber \\
&& {} \Big. \sum_{i=1}^K \frac{d_i}{\min(N_i, M_{\rm{tot}}) } \leq 1
\Big\}. \label{eq: Kuser CRC outer-bound}
\end{eqnarray}
\end{lemma}
\begin{IEEEproof}
Since every transmitter knows the message of the primary, it follows
$d_1 \leq \min(M_{\rm{tot}}, N_1)$. Also we have the single-user bounds
$d_i \leq \min(M_i, N_i)$ $\forall$ $i>1$. The bound on the weighted
sum of $\{d_i\}$'s holds because the overall BC outer-bound is
applicable to the CRC as well.
\end{IEEEproof}

The outer-bound is tight in the following cases.
\begin{theorem} \label{thm: K-user CRC}
The DoF region of the $K$-user MIMO CRC with no CSIT is given by the
region defined in equation (\ref{eq: Kuser CRC outer-bound}),
provided one of the following conditions hold:
\begin{enumerate}
\item $N_i \leq M_i$ $\forall$ $i>1$, and
\item $N_i = N$, $M_i = M$ $\forall$ $i$ and $N > M$.
\end{enumerate}
\end{theorem}
\begin{IEEEproof}
Again, only the achievability part needs to be proved. \newline Case
1): The outer-bound in this case is defined by the constraint $
\frac{d_1}{\min(N_1, M_{\rm{tot}})} + \sum_{i=2}^K
\frac{d_i}{\min(M_i, N_i)} \leq 1$. This entire region is achievable
by time division. \newline Case 2): In this case, the outer-bound is
defined by the constraints $d_i \leq M$ $\forall$ $i>1$ and $\sum_i
d_i \leq \min(KM,N)$. Since $\sum_i d_i \leq N$, the receivers can
simply zero-force the interference. Now, it must be verified that
the transmitters can support any DoF-tuple in the region. Given any
$K$-tuple $(d_1,d_2,\cdots,d_K)$, the $i^{th}$ transmitter, $i>1$, must
send $d_i$ $(\leq M)$ streams since no other transmitter is
cognitive of its message. Therefore, all the transmitters
together can send $M + \sum_{i=2}^K (M-d_i)$ streams to the first
receiver since every transmitter knows the message of the primary
pair. For any $K$-tuple $(d_1, \cdots, d_K)$ with $\sum_i d_i \leq KM$,
$d_1 \leq KM - \sum_{i=2}^K d_i = M + \sum_{i=2}^K (M-d_i)$, and
hence, the transmitters can support any $K$-tuple in the outer bound.
Therefore, the entire outer-bound is also achievable.
\end{IEEEproof}

\begin{remark}
Note that the above theorem provides the complete characterization
in the case wherein all transmitters as well as all receivers have
equal number of antennas.
\end{remark}

\subsection{The MIMO X Channel}

\begin{figure} \centering
\includegraphics[height=3.5in,width=4.5in]{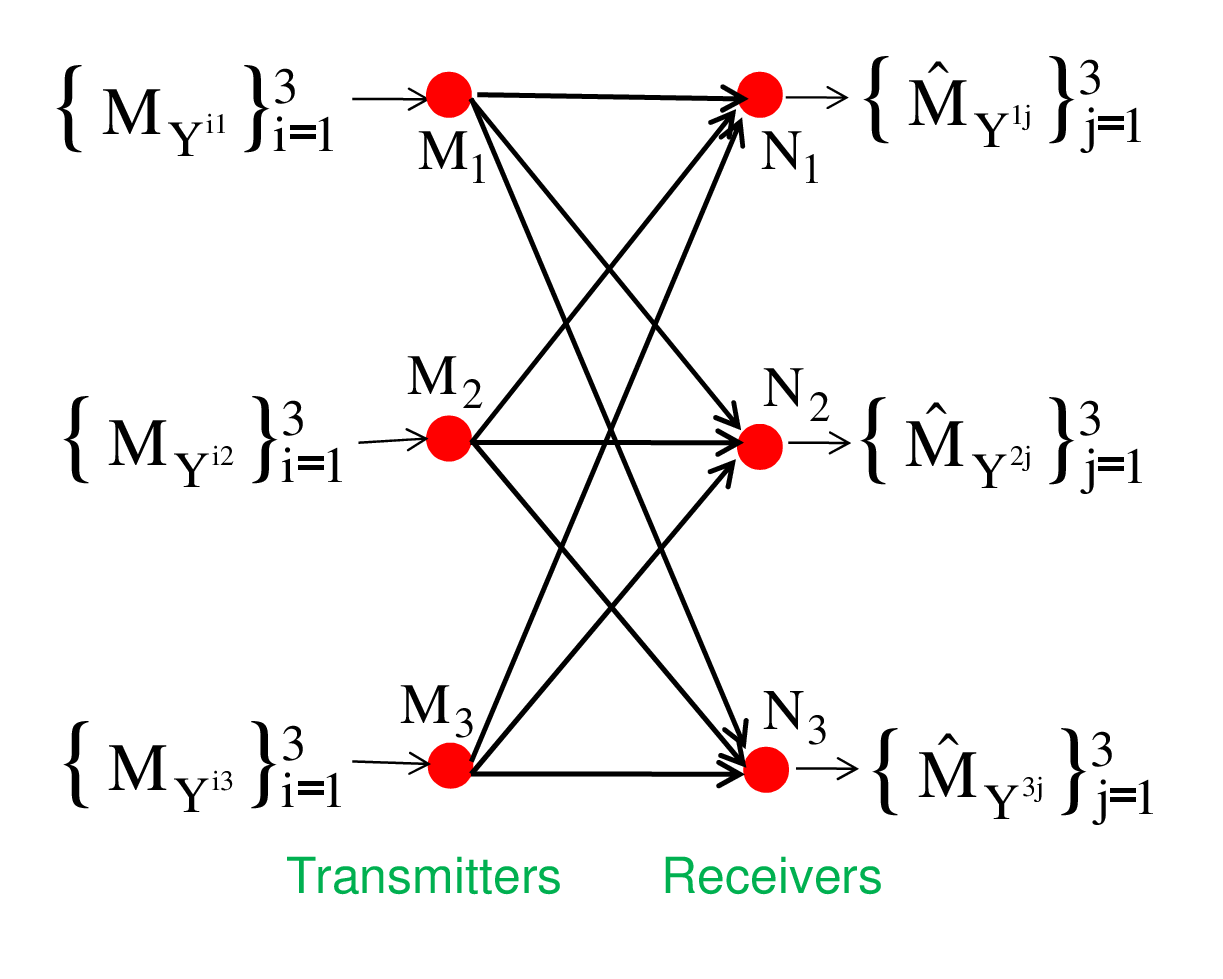}
\caption{The $K$-User MIMO X Channel with $K = 3$.} \label{fig:
MIMO_Kuser_Xchannel}
\end{figure}

The $K$-user X channel is like the fully connected $K$-user IC, except that
every transmitter has a message for every receiver. Fig.
\ref{fig: MIMO_Kuser_Xchannel} shows an example of the
$K$-user X channel with $K = 3$. Let $d_{ji}$ denote the DoF
corresponding to the message sent by the $i^{th}$ transmitter to the
$j^{th}$ receiver. Then we have by the following theorem.

\begin{theorem} \label{thm: dof X channel}
The DoF region of the $K$-user no-CSIT MIMO X channel with $M_i = M$
and $N_i = N$ $\forall$ $i$ is given by
\[
\mathbf{D} = \left\{ \Big( d_{ij} \Big)_{i,j=1}^K \left| 0 \leq
d_{ij} \leq \min(M,N) ~ \forall ~i,j, ~ \sum_{i,j = 1}^K d_{ij} \leq
\min(N,KM) \right. \right\} .
\]
\end{theorem}
\begin{IEEEproof}
Let us first prove that the above region is an outer-bound. Clearly,
$d_{ij} \leq \min(M,N)$ $\forall$ $i$, $j$. To prove the remaining
inequality, let use assume that all the transmitters can cooperate.
Then Theorem \ref{thm: dof region BC} implies that the following
inequality is a valid outer-bound:
\[
\sum_{i=1}^K \frac{\sum_j d_{ij}}{\min(N,KM)} \leq 1 \Rightarrow
\sum_{i,j=1}^K d_{ij} \leq \min(N,KM).
\]
The fact that the above region is an inner-bound follows directly
from Theorem \ref{thm: K-user IC}.
\end{IEEEproof}

\begin{remark}
The above theorem can be easily generalized to the case of $X$
channel with unequal number of transmitters and receivers. The
generalization has been omitted to avoid repetition.
\end{remark}

\begin{remark}
The inner and outer-bounds to the DoF region of the perfect-CSIT
$2$-user X channel have been proposed and these bounds are known to
be tight in the special case wherein all terminals have equal number
of antennas \cite{JaferShamai, Khandani_Xchannel_constant}. This
work was generalized in \cite{Jafar_Cadambe_wireless_X_network}
to the case of time-varying X network with an unequal number of
transmitters and receivers (both $\geq 2$). In particular,
\cite{Jafar_Cadambe_wireless_X_network} characterizes
the exact DoF region of the SISO X channel. However, for the case
wherein all terminals have multiple but equal number of antennas,
only inner and outer-bounds are given which do not coincide.
We hence have an exact characterization of the no-CSIT DoF
region although the perfect-CSIT DoF region is not yet completely known.
\end{remark}

\section{Conclusion}
In this paper, we comprehensively deal with the problems of obtaining the DoF regions
without CSIT of several MIMO networks including broadcast, interference, X and cognitive radio networks
for the 2-user case and $K$-user cases under general conditions on the fading distributions that
subsume the commonly assumed models of i.i.d. Rayleigh fading and isotropic fading as special cases.
The exact characterization of the DoF region of the $K$-user BC is obtained. For the two-user
MIMO IC and the CRC with an arbitrary numbers of antennas at each node, the inner and outer bounds
obtained herein yield the exact characterization of the DoF regions, except for a few cases. Finally, the DoF
regions of some important classes of the $K$-user MIMO IC, X,
CRC and multi-hop interference networks are also derived. Comparisons with perfect CSIT DoF regions in many cases
reveal insights about when  a lack of CSIT results in a loss of DoF, thereby motivating feedback of CSI in these cases and when lack of CSIT results in no loss of DoF yielding robust CSI-independent transmission schemes in these cases
that achieve the DoF performance of their CSI-dependent counterparts resulting from the corresponding previous perfect CSIT study. An interesting open problem is to find the DoF region of the CRC for the case in which it is not unknown. More broadly speaking, the results derived here for the no-CSIT case warrant a generalization to the case of partial CSIT.

\appendices
\section{The Multi-Hop Interference Network} \label{app: two_hop_IC}

\begin{figure} \centering
\includegraphics[height=2.5in,width=5in]{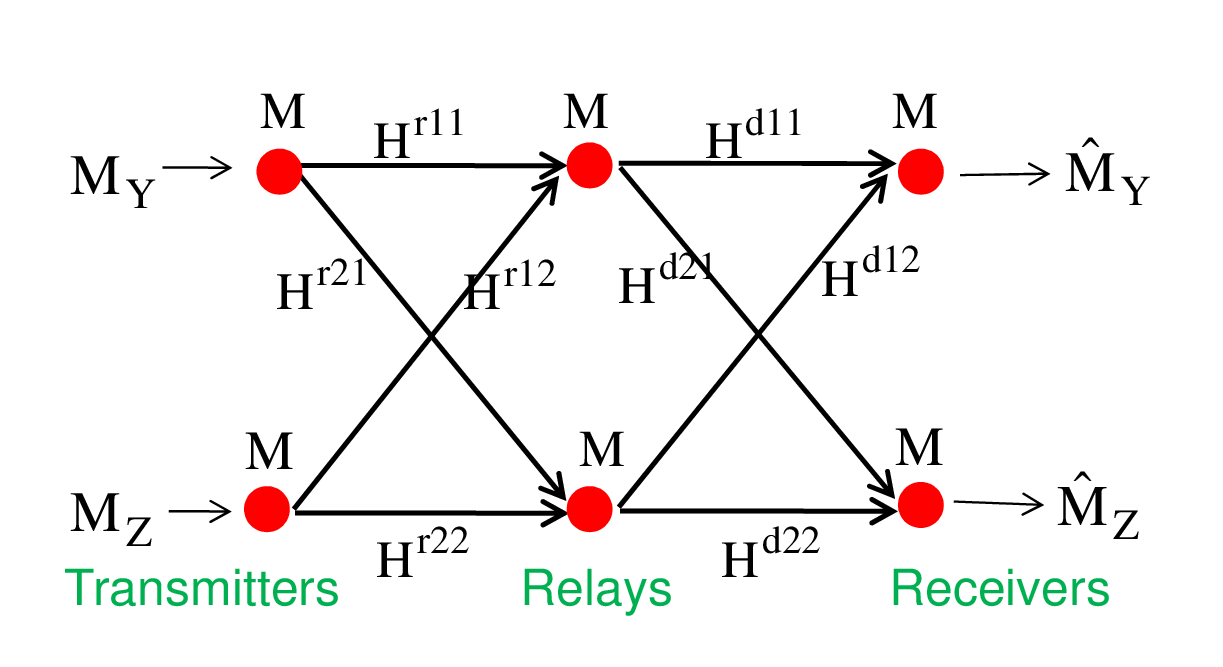}
\caption{The $2$-Hop $2$-User Interference Channel (IC).}
\label{fig: 2hop_2user_IC}
\end{figure}

In this appendix we obtain the DoF region of a general
multi-hop ($n$-hop) interference network with $K$ transmit-receive terminal pairs with
$n-1$ layers of $K$ relays each that separate them, wherein the last layer of relays
has no knowledge of the outgoing channel realizations to the receivers. All terminals are
assumed to have $M$ antennas each. It is sufficient to consider the particular case of the
two-user, two-hop IC because the argument easily extends to the general case.

The two-user, two-hop IC consists
of two transmitters which intend to communicate two independent
messages to their respective receivers and this communication is
aided by two relays as shown in Fig. \ref{fig: 2hop_2user_IC}. The
input-output relationship is defined via the equations
\begin{eqnarray*}
Y^{r1}(t) = H^{r11}(t) X^{1}(t) + H^{r12}(t) X^{2}(t) + W^{r1}(t), \\
Y^{r2}(t) = H^{r21}(t) X^{1}(t) + H^{r22}(t) X^{2}(t) + W^{r2}(t), \\
Y^{d1}(t) = H^{d11}(t) X^{r1}(t) + H^{d12}X^{r2}(t) + W^{d1}(t), \\
Y^{d2}(t) = H^{d21}(t) X^{r1}(t) + H^{d22}X^{r2}(t) + W^{d2}(t),
\end{eqnarray*}
where at the $t^{th}$ time slot, $X^1(t)$, $X^2(t)$, $X^{r1}(t)$,
and $X^{r2}(t) \in \mathbb{C}^{M \times 1}$ are the signals
transmitted by the two transmitters and the two relays,
respectively, $Y^{r1}(t), \; Y^{r2}(t)$ and $Y^{d1}(t), \; Y^{d2}(t)
\in \mathbb{C}^{M \times 1}$ are the signals received by the two
relays and the two receivers, respectively; and $W^{r1}(t)$,
$W^{r2}(t)$, $W^{d1}(t)$, and $W^{d2}(t) \in \mathbb{C}^{M \times
1}$ are the additive white Gaussian noises. This channel has been studied recently under the assumption of
perfect and global CSIT (see \cite{Cao_Chen_two_hop_IC_2009,Jafar_2hop_IC_DoF} and the references therein).

Define $H^{d1}(t) = \begin{bmatrix} H^{d11}(t) & H^{d12}(t)
\end{bmatrix}$ and $H^{d2}(t) = \begin{bmatrix} H^{d21}(t) &
H^{d22}(t) \end{bmatrix}$. The following theorem gives the DoF
region of this channel when CSI is not available at the relays.
\begin{theorem}
Consider the two-hop IC with AWGN in which $H^{d1}$ and $H^{d2}$
follow a distribution of type $\mathcal{D}_0(2M,M,M)$. Suppose that both
receivers know the channel matrices $H^{d1}(t)$ and $H^{d2}(t)$
perfectly and instantaneously, whereas the relays know only their
distribution. The DoF region of this two-hop IC is then
given by
\[
\mathbf{D} = \Big\{ (d_1,d_2) \left| d_1, d_2 \geq 0, ~ d_1 + d_2
\leq M \right. \Big\} ,
\]
regardless of the knowledge of the channel matrices $\{
H^{rij}(t)\}_{i,j=1}^2$ at the transmitters, relays, and the
receivers.
\end{theorem}
\begin{IEEEproof}
The achievability of the above region follows by time division. To
prove the converse, consider the following argument. The DoF region
of the two-hop IC can not reduce if both relays are given the
side information of the two messages. Hence, the DoF region of the
two-hop IC is outer-bounded by that of the BC in which the two
relays serve as the common transmitter (since both of them know both
messages) and the receivers of the two-hop IC are
also the receivers in the BC. Hence, using Theorem \ref{thm: dof region BC},
we have that for any $(d_1,d_2) \in \mathbf{D}$, $d_1 + d_2 \leq M$.
\end{IEEEproof}

\begin{remark}
The above theorem shows that even if the transmitters have perfect
CSI, it can not help improve the DoF region unless the relays have
CSI of $H^{d1}(t)$ and $H^{d2}(t)$.
\end{remark}

\begin{remark}
Recently, it has been proved that the single-antenna $2$-hop IC (with $M=1$)
has $2$ sum-DoF, if there is perfect CSI at all terminals
\cite{Jafar_2hop_IC_DoF} (see also references therein). The above
theorem shows that the achievability of 2 DoF over this channel depends
critically on having perfect knowledge of $H^{d1}(t)$ and
$H^{d2}(t)$ at the relays.
\end{remark}

This result can clearly be extended to the $n$-hop $K$-user interference network
consisting of $K$ transmitters, which need to communicate $K$
independent messages to their respective receivers through $(n-1)$
layers of relays where, at each layer, there are $K$ relays and the
relays of the last layer do not have knowledge of the channel
matrices of the last hop.

\section{Proof of Theorem \ref{thm:D1}} \label{app: proof of thm:D1}

Let us first consider the case of AWGN.
To prove that the outer-bound is still given by the region defined
in equation (\ref{eq: dof region BC}), it is sufficient to prove
that Step II.b follows under the generalization of fading distribution being in
$\mathcal{D}_1 (M,\bar{N})$. To this end, note
that the property of statistical equivalence of random variables
$\mathbf{\tilde{Y}^{i-1}}_1$, $\mathbf{\tilde{Y}^{i-1}}_2$,
$\cdots$, $\mathbf{\tilde{Y}^{i-1}}_{N_{i-1}}$, and
$\mathbf{\tilde{Y}^i}_1$, $\mathbf{\tilde{Y}^i}_2$, $\cdots$,
$\mathbf{\tilde{Y}^i}_{N_i}$ still holds. Hence, Step II.b follows
and the conclusion of Theorem \ref{thm: dof region BC} applies.

Next consider the case of ACGN but with fading distribution in
$\mathcal{D}_0 (M,\bar{N})$. Let $d_{\min}$ be the minimum of all eigenvalues of all covariance
matrices $\{\Sigma^i\}_{i=1}^K$. Since $\{\Sigma^i\}$'s are taken to
be positive definite, $d_{\min} > 0$. Then $\Sigma_i \succeq
d_{\min} I_{N_i} ~ \forall ~i$ within the partial order of positive
semi-definite matrices. Therefore, if we assume that $W^i(t) \sim
\mathcal{C}\mathcal{N}(0,d_{\min} I_{N_i})$, then the technique of
channel enhancement developed earlier implies that the involved MI
terms can only increase by this assumption. Therefore, the DoF
region of the BC with $W^i(t) \sim \mathcal{C}\mathcal{N}(0,d_{\min}
I_{N_i})$, which is given by Theorem \ref{thm: dof region BC}, will
be an outer-bound to the DoF region of the original BC where $W^i(t)
\sim \mathcal{C}\mathcal{N}(0,\Sigma_i)$. The result then follows by
noting that the region defined in Theorem \ref{thm: dof region BC}
is achievable over the original BC.

The general case with fading distribution in
$\mathcal{D}_1 (M,\bar{N})$ and with ACGN is now easily proved by combining the two arguments above.

\section{The Two-User MIMO IC is Separable in the DoF
Sense} \label{app: IC separable dof sense}

It is shown here that the perfect-CSIT DoF region of the IC with
i.i.d. fast fading is given by the region defined in equation
(\ref{eq: dof region IC perfect}). Recall that the DoF region of the
(perfect-CSIT) IC with the deterministic channel fading matrices is
derived in \cite{Jafar-Maralle}. Their proof is suitably modified
here to obtain the result. The proof here makes use of the fact that
the point-to-point perfect-CSIT MIMO channel is separable
\cite{Cadambe_Jafar_separability, Caire_Shamai_channels_with_SI}.

From equation (\ref{eq: dof region IC perfect}), we see the
sufficiency of proving $d_1 + d_2 \leq \max(N_1,M_2)$, which is the
topic of the remainder of the appendix. It is first shown that if
$N_1 \geq M_2$, then $d_1 + d_2 \leq N_1$ (c.f. \cite[Theorem
1]{Jafar-Maralle}), and later the remaining case is dealt with (c.f.
\cite[Corollary 1]{Jafar-Maralle}).

Let us start with the case of $N_1 \geq M_2$. In the notations of
Section \ref{subsec: IC outer-bound proof}, we have, by Fano's
inequality,
\[
R_1  \leq \frac{1}{n} I(M_Y;\mathbf{Y} ,\mathbf{H}) + \epsilon_n
\mbox{ and } R_2  \leq \frac{1}{n} I(M_Z;\mathbf{Z} , \mathbf{H}) +
\epsilon_n .
\]
Now, define
\[
\alpha(t) = \min \Big( \frac{1}{\sigma^2_{\max}[H^{12}(t)]},
\frac{1}{\sigma^2_{\max}[H^{22}(t)]} \Big),
\]
where $\sigma_{\max}[A]$ represents the largest singular-value of
$A$; and
\[
W_b(t) \sim \mathcal{C}\mathcal{N} \left( 0, H^{12}(t) \Big\{
[H^{12}(t)]^* H^{12}(t) \Big\}^{-1} [H^{12}(t)]^* - \alpha(t)
H^{12}(t) [H^{12}(t)]^* \right)
\]
and the realizations of $W_b(t)$ are i.i.d. across time. Then
following Steps 1 and 2 of the proof of \cite[Theorem
1]{Jafar-Maralle}, we get the following:
\begin{eqnarray*}
R_1 & \leq & \frac{1}{n} I \big( M_Y; \mathbf{Y}- \mathbf{W}_b |
\mathbf{H} \big) + \epsilon_n \\
& = & \frac{1}{n} I \big( M_Y,M_Z; \mathbf{Y}- \mathbf{W}_b |
\mathbf{H} \big) + \epsilon_n - \frac{1}{n}
I \big( M_Z; \mathbf{Y}- \mathbf{W}_b | M_Y, \mathbf{H} \big) + \epsilon_n, \mbox{ and} \\
R_2 & \leq & \frac{1}{n} I(M_Z ; \mathbf{Z}  | M_Y, \mathbf{H} \big)
+ \epsilon_n.
\end{eqnarray*}
Then following Steps 3-5 of the proof of \cite[Theorem
1]{Jafar-Maralle}, it can be shown that
\[
I \big( M_Z; \mathbf{Y}- \mathbf{W}_b | M_Y, \mathbf{H} \big) \geq
I(M_Z ; \mathbf{Z}  | M_Y, \mathbf{H} \big).
\]
This implies that
\[
R_1 + R_2 \leq \frac{1}{n} I \big( M_Y,M_Z; \mathbf{Y}- \mathbf{W}_b
| \mathbf{H} \big) + \epsilon_n.
\]
In other words, $R_1 + R_2$ must be less than or equal to the rate
achievable over some perfect-CSIT point-to-point MIMO channel with
$M_1 + M_2$ transmit and $N$ receive antennas. The total DoF
achievable over such a channel are always limited by $N_1$ since it
is separable \cite{Telatar, Caire_Shamai_channels_with_SI,
Zamir_Erez_power_allocation}, \cite[Lemma 6]{Lapidoth}. Hence, $d_1
+ d_2 \leq N_1$ as needed.

Further, using \cite[Corollary 1]{Jafar-Maralle}, the remaining case
of $M_2 > N_1$ can also be handled to prove the required bound on
the sum-DoF.

It is also easy to see that the assumption of fading matrices
varying independently across time is not critical to the above
proof. We only need that the fading process to be stationary and
ergodic.

\section{Proof of Theorem \ref{thm: IC generalization}} \label{app: proof of thm: IC generalization}

We prove this theorem along the lines of the proof
of Theorem \ref{thm: outer bound IC}, which, as one may recall,
consists of three steps. Moreover, the last step, Step III,
is insensitive to the distribution of channel matrices and thus
follows without any modification. Therefore, we focus on Steps I
and II. Toward this end, for each $j$, we redefine $\tilde{Y}(t)$
and $\tilde{Z}(t)$ (which would serve as counterparts of equations
(\ref{eq:IC channel enhanced outputs1}) and
(\ref{eq:IC channel enhanced outputs2}) in the proof of Theorem
\ref{thm: outer bound IC}) such that the inequalities in (\ref{eq:
channel enhancement IC}) (and hence, in (\ref{eq: IC bounds after
Step I})) and (\ref{eq: IC main inequality}) still hold, even though
the channel matrices follow a distribution of type $\mathcal{D}_j$.
Further, the validity of inequalities (\ref{eq: channel enhancement
IC}) and (\ref{eq: IC bounds after Step I}) implies that of Step I,
whereas the implication of the inequality in (\ref{eq: IC main
inequality}) is that Step II holds. The proof of this theorem can
then be completed by performing Step III.

Consider the case of $j = 1$, where the channel
matrices follow a distribution of type $\mathcal{D}_1$. Here,
$\tilde{Y}(t)$ and $\tilde{Z}(t)$ are defined in a manner identical
to their definitions stated in the proof of Theorem \ref{thm: outer
bound IC}. Then, the arguments which yield us inequalities (\ref{eq:
channel enhancement IC}) and (\ref{eq: IC bounds after Step I}) in
the case of Theorem \ref{thm: outer bound IC} imply their validity
in the present case as well. It now remains to verify that the
inequality in (\ref{eq: IC main inequality}) holds. To this end,
recall that this inequality has been stated before as Lemma
\ref{lem: IC main inequality}, and moreover, as argued in its proof
there, this inequality holds, provided we have the property of
statistical equivalence of $\mathbf{\tilde{Y}}$ and
$\mathbf{\tilde{Z}}$, which can be shown to be true, even with the
channel matrices of type $\mathcal{D}_1$ (c.f. the proof of Theorem
\ref{thm:D1}, which generalizes the result of Theorem \ref{thm: dof
region BC} from class $\mathcal{D}_0$ to class $\mathcal{D}_1$).
Hence, the theorem follows for $j = 1$.

When $j = 2$, we define $\tilde{Y}(t)$ and
$\tilde{Z}(t)$ as follows. Write $H^{i1}(t) = U^{i1}(t)
\Lambda^{i1}(t) \big( V^{i1}(t) \big)^*$ for $i = 1,~2$ (see
Definition \ref{def: thm 2.5}). Let $h(t)$ to be the maximum of all
elements of matrices $\Lambda^{11}(t)$ and $\Lambda^{21}(t)$. Define
$h^1_{\max}(t) = \max\{ 1, h(t)\}$ and $h^{i1}_{\max}(t)
=h^1_{\max}(t) I_{N_i \times N_i}$ for $i=1,~2$. Let
$\tilde{\Lambda}^{i1}(t)$ be the square diagonal matrix formed by
taking only the first $\min(M_1,N_i)$ rows of $\Lambda^{i1}(t)$
(i.e., it contains singular values of $H^{i1}(t)$ along the
diagonal). Define the matrices
\[
D^{i1}(t) = \begin{bmatrix} \tilde{\Lambda}^{i1}(t) & 0_{p_i \times
q_i} \\
0_{q_i \times p_i} & I_{q_i \times q_i} \end{bmatrix} \mbox{ for } i
=1,~2,
\]
where $p_i = \min(M_1,N_i)$, $q_i = N_i - p_i$, and $0_{p_i \times
q_i}$ denotes the all-zero matrix of size $p_i \times q_i$. Then
define
\begin{eqnarray*}
\tilde{Y}(t) & = & h^{11}_{\max}(t) \big( D^{11}(t) \big)^{-1}
\big( U^{11}(t) \big)^* H^1(t) \begin{bmatrix} X^1(t) \\
X^2(t) \end{bmatrix} + W(t) \mbox{ and } \\
\tilde{Z}(t) & = & h^{21}_{\max}(t) \big( D^{21}(t) \big)^{-1}
\big( U^{21}(t) \big)^* H^2(t) \begin{bmatrix} X^1(t) \\
X^2(t) \end{bmatrix} + W'(t).
\end{eqnarray*}
With the above definitions, the inequalities in (\ref{eq: channel
enhancement IC}) and (\ref{eq: IC bounds after Step I}) are true
because all the diagonal elements of matrices $D^{11}(t)$ and
$D^{21}(t)$ are less than or equal to $h^1_{\max}(t)$, which is
present along the diagonal of $h^{11}_{\max}(t)$ and
$h^{21}_{\max}(t)$. Further, the property of statistical equivalence
of $\mathbf{\tilde{Y}}$ and $\mathbf{\tilde{Z}}$ also holds (see
Definition \ref{def: thm 2.5}) and hence also the inequality
(\ref{eq: IC main inequality}). The theorem then follows for $j = 2$
by completing Step III.

\bibliographystyle{IEEEtran}
\bibliography{v12_DoF_journal.bbl}
\end{document}